\documentclass[11pt,preprint]{aastex}

\def\simlt{\lower.5ex\hbox{$\; \buildrel < \over \sim \;$}}

\usepackage{graphicx}
\usepackage{natbib}
\usepackage{epstopdf}
\usepackage{epsfig}
\usepackage{subfig}
\usepackage{color}
\usepackage{datetime}
\usepackage{bbding}

\begin{document}

\title{A VLA Survey of Radio-Selected SDSS Broad Absorption Line Quasars}
\author{M. A. DiPompeo\altaffilmark{1}, M. S. Brotherton\altaffilmark{1}, C. De Breuck\altaffilmark{2}, Sally Laurent-Muehleisen\altaffilmark{3}}
\altaffiltext{1}{University of Wyoming, Dept. of Physics and Astronomy 3905, 1000 E. University, Laramie, WY 82071, USA}
\altaffiltext{2}{European Southern Observatory, Karl Schwarzschild Strasse 2, 85748 Garching bei M\"{u}nchen, Germany}
\altaffiltext{3}{Illinois Institute of Technology, 3101 South Dearborn St., Chicago, IL 60616, USA}

\begin{abstract}
We have built a sample of 74 radio-selected broad absorption line quasars from the Sloan Digital Sky Survey Data Release 5 (SDSS DR5) and Faint Images of the Radio Sky at Twenty Centimeters (FIRST), along with a well matched sample of 74 unabsorbed ``normal'' quasars.  The sources have been observed with the NRAO Very Large Array/Expanded Very Large Array at 8.4 GHz (3.5 cm) and 4.9 GHz (6 cm).  All sources have additional archival 1.4 GHz (21 cm) data.  Here we present the measured radio fluxes, spectral indices, and our initial findings.  The percentage of BAL quasars with extended structure (on the order of 10\%) in our sample is similar to previous studies at similar resolutions, suggesting that BAL quasars are indeed generally compact, at least at arsecond resolutions.  The majority of sources do not appear to be significantly variable at 1.4 GHz, but we find two previously unidentified BAL quasars that may fit into the ``polar'' BAL category.  We also identify a significant favoring of steeper radio spectral index for  BAL compared to non-BAL quasars.  This difference is apparent for several different measures of the spectral index, and persists even when restricting the samples to only include compact objects.  Because radio spectral index is a statistical indicator of viewing angle for large samples, these results suggest that BAL quasars do have a range of orientations but are more often observed farther from the jet axis compared to normal quasars.

\end{abstract}


\section{INTRODUCTION}
Radio observations of broad absorption line (BAL) quasars have been interpreted to indicate the inability of simple orientation schemes to explain the observed incidence of BALs in quasar spectra.  The simplest orientation models suggest that all quasars have outflows, but only approximately 20\% (the observed fraction of optically selected quasars which exhibit BALs, e.g. Knigge et al. 2008) are seen from a line of sight that intercept these outflows (Weymann et al 1991).  In this model BAL winds are radiatively accelerated from the edge of the dusty torus surrounding the quasar central engine or from the accretion disk itself; i.e. BAL quasars are seem from a more edge-on perspective, and unabsorbed quasars are seen more face-on.  Around one third of BAL quasars also show a relatively high level of optical polarization (Hines \& Schmidt 1997; DiPompeo et al. 2010; DiPompeo et al. 2011), and this has been historically taken to support the simple orientation model (eg. Goodrich \& Miller 1995; Hines \& Wills 1995).

Brotherton et al. (1998) were among the first to identify radio-loud BAL quasars, surprising because they were once thought to be strictly radio quiet (Stocke et al. 1992).  This was made possible mostly due to large radio surveys such as FIRST (Becker et al. 1995) and NVSS (Condon et al. 1998), and allowed the first true studies of the orientation of BAL quasars.  Although BAL quasars often do not have extended radio structure which would allow a direct estimate of their orientation, the steepness of radio spectrum can indicate orientation (Fine et al. 2011).  In general, more edge-on radio sources are dominated by optically thin lobe emission, which has a steep spectrum, and more face-on sources are dominated by relativistically boosted core radio emission, which is optically thick and thus flatter due to synchrotron self absorption.  Studies so far have shown that BAL quasars have a wide range of radio spectral indices ($\alpha$, defined as $f \propto \nu^{\alpha}$, where $f$ is the radio flux and $\nu$ is the frequency), suggesting a full range of orientations (Becker et al. 2000).  A significant difference between the distribution of $\alpha$ for BAL and non-BAL quasars has not been previously identified, but the sample sizes have typically been small (Fine et al. 2011).  Montenegro-Montes et al. (2008) did identify a difference in the $\alpha$ distributions of their samples- however when they restricted their analysis to include only compact sources the difference disappeared.  Additionally, BAL quasars with polar rather than equatorial outflows have likely been identified via short timescale variation in radio flux (Zhou et al. 2006; Ghosh \& Punsly 2007).

The existence of apparently polar BALs and the fact that BAL quasars are often compact in radio maps suggests a different picture.  Becker et al. (2000) find that only about 10\% of BAL quasars show extended structure at 5\arcsec\ resolution, compared to about 50\% of normal quasars.  In a sample of 15 BAL quasars, Montenegro-Montes et al. (2008) find that all of them are compact at FIRST resolutions, and the majority remain compact at around 80 mas resolution.  Even very long baseline interferometry (VLBI) observations, which has down to milli-arcsecond resolution, often show BAL quasars as compact objects (Doi et al. 2009, Jiang \& Wang 2008, Kunert-Bajraszewska et al. 2009).  The fact that it is required to observe many of these objects on size scales on the order of a few hundred parsecs or less before seeing any resolved structure (Kunert-Bajraszewska et al. 2010, Liu et al. 2008) suggests that they are intrinsically quite small.  This size scale along with a convex spectral shape (which has been seen in many BAL quasars; e.g. Montenegro-Montes et al. 2008) is typical of the compact steep spectrum (CSS) and gigahertz-peaked spectrum (GPS) sources. Objects of this type are thought to be young radio sources (O'Dea 1998, Fanti et al 1995).  Instead of having a preferred line of sight, maybe there is a BAL phase in all quasars which lasts for approximately 20\% of the quasar lifetime.  A model where a BAL phase evolves into a radio-loud phase, with a relatively short overlap has been proposed (Gregg et al. 2002, 2006).  

The studies of radio-selected BAL quasars have thus far been with small samples.  In order to find more concrete results a large, statistically significant sample with radio data at multiple frequencies is needed.  This project aims at nailing down the radio properties of BAL quasars conclusively to finally find a consistent model for these important objects.  Here we will present the results of our NRAO\footnote{The National Radio Astronomy Observatory is a facility of the National Science Foundation operated under cooperative agreement by Associated Universities, Inc.} Very Large Array/Expanded Very Large Array (VLA/EVLA) observations.

We adopt the cosmology of Spergel et al. (2007) for all calculated properties, with $H_0 = 71$ km s$^{-1}$ Mpc$^{-1}$, $\Omega_M=0.27$ and $\Omega_{\Lambda} =0.73$.

\section{TARGETS}
In order to build a large sample, we started with the BAL quasar catalog of Gibson et al. (2009), which is drawn from the Sloan Digital Sky Survey Data Release 5 (SDSS DR5).  Gibson et al. (2009) use both the traditional definition of ``balnicity index'' ($BI$; Weymann et al. 1991) and a modified $BI$ ($BI_0$, which integrates absorption starting from 0 km s$^{-1}$ instead of the traditional $-$3000 km s$^{-1}$) in the lines of \ion{C}{4}, \ion{Si}{4}, \ion{Al}{3}, and \ion{Mg}{2} to identify BAL quasars.  They also require absorption to be continuous over a range of 2000 km s$^{-1}$ (see their paper for additional details).  

In order to pick out radio bright sources, we matched the optical positions from the Gibson et al. (2009) sources to both the FIRST and NVSS catalogs, selecting sources that matched within 10\arcsec\ of FIRST radio positions and had integrated fluxes in either survey of more than 10 mJy.  We independently matched to both catalogs due to the differences in resolution of each; FIRST has a resolution of 5\arcsec, while NVSS is much lower at 45\arcsec.  We wanted the higher sensitivity of FIRST, but only matching with the FIRST catalog may miss objects with faint core emission but bright extended lobe emission.  Also matching to NVSS allowed us to identify possible extended sources, investigate the FIRST maps to verify that the sources were indeed extended, and add up the flux in individual FIRST components to get a total source integrated flux.  Thus we put together a BAL quasar sample which had $\ge$10 mJy FIRST fluxes, including those with extended, individually resolved components.  As discussed below, many observations were to be carried out in the VLA D-array, which has the smallest baseline and thus lowest resolution.  Therefore we inspected the FIRST maps of all sources by eye to eliminate any in which contamination of nearby sources may be an issue.
  
We also applied a redshift cut of $z \ge 1.5$ in order to include the \ion{C}{4} emission line in the spectral window of SDSS.  Each spectrum remaining at this point was inspected by eye to ensure that all objects were indeed BAL quasars.  Gibson et al. (2009) also examined their spectra by eye, but we wanted to make sure we agreed that all sources were unambiguously BAL quasars; while no strict signal-to-noise cut was applied, some of the SDSS spectra were too noisy for us agree with certainty that features were BALs and so a few sources were thrown out.  

In the end, we were left with a sample of 74 radio-selected BAL quasars.  Table~\ref{balpropstbl} lists the sample and its properties: column (1) is the SDSS source name, (2) is the BAL quasar subtype (Hi$=$high-ionization BAL, Lo$=$low-ionization BAL, Fe$=$Iron LoBAL), (3) and (4) are the RA and DEC of the source at 1.4 GHz (from FIRST), (5) is the SDSS redshift, (6) is the absolute $B$ magnitude, and (7) is the FIRST integrated 1.4 GHz flux (sum of individual components if necessary).  The remaining columns will be discussed below.

In order to make meaningful comparisons with the parent populations of quasars, a well matched sample of normal, unabsorbed quasars was developed.  For each of the individual BAL quasars in our sample, we searched the SDSS database for a normal quasar that matched within 20\% SDSS i-band magnitude, 20\% of 1.4 GHz radio flux, and 10\% of redshift.  Each quasar spectrum was examined to ensure they were unabsorbed, and the FIRST maps were inspected for extended structure and possible contaminating sources.  Thus, a one-to-one matched sample of 74 non-BAL quasars was built for use as a comparison.  Table~\ref{nbalpropstbl} lists the properties of the non-BAL sample; columns are the same as Table~\ref{balpropstbl} with the subtype column omitted, since all are normal quasars.  

\begin{figure}
 \centering
  \figurenum{1}
   \includegraphics[width=4in,height=5in]{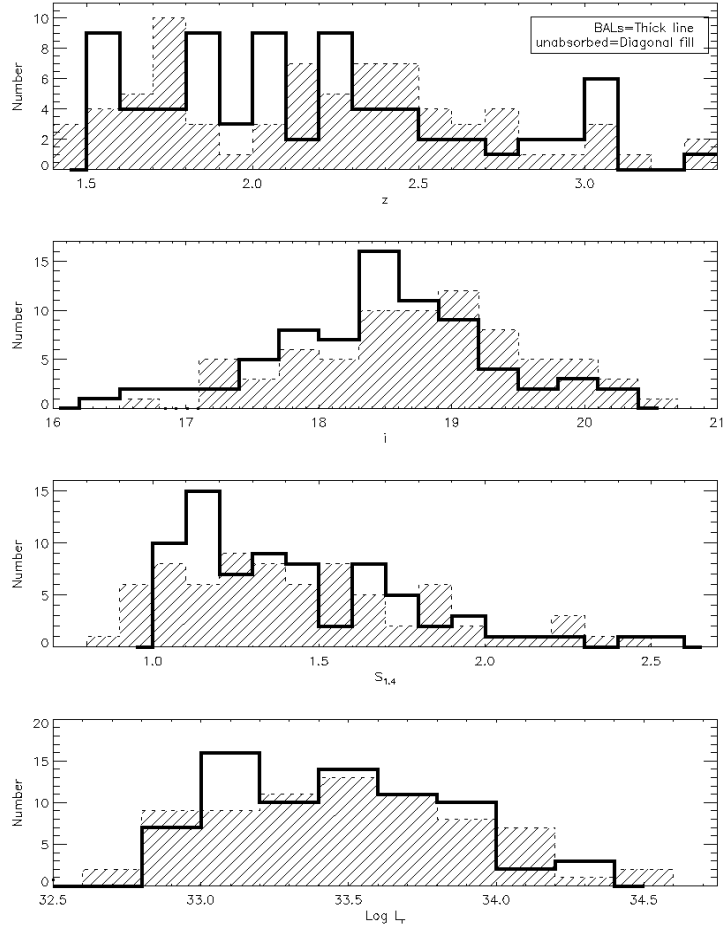}
  \caption{Comparison of the properties used to build the matched sample of non-BAL quasars; SDSS i-band ($i$), redshift ($z$), and FIRST integrated flux ($S_i$).  Because each individual BAL source was matched within a percentage of these properties to a non-BAL source, the histograms are not identical but similar.  The final panel shows the distributions of radio luminosity, with a k-correction to rest frame 4.9 GHz, to confirm that matching in redshift and flux indeed provided a sample well matched in intrinsic luminosity.  K-S tests on the distributions do not indicate significant differences.\label{comparisonfig}}
\end{figure}

The first 3 panels of Figure~\ref{comparisonfig} show a comparison of the properties that the samples were matched on.  The final panel of this figure shows the K-corrected rest-frame 4.9 GHz luminosity distributions of the samples, to illustrate that the sample matching in flux and redshift did indeed result in a matched luminosity.  The luminosities shown in this figure used the integrated FIRST flux at 1.4 GHz and radio spectral index between 1.4 and 4.9 GHz (as discussed below) for the K-correction.

\section{OBSERVATIONS \& MEASUREMENTS}
\subsection{Observations}
Observations were performed over two observing periods, all with dynamically scheduled time, at two frequencies: 8.4 GHz (3.5 cm, X-band) and 4.9 GHz (6 cm, C-band).  The ultimate aim was to measure the radio spectral index for all sources.  Individual objects were observed at both frequencies in the same observing block (with a few exceptions, see below), so the two measurements were essentially simultaneous.  Since a significant number of BAL quasars are compact at VLA resolutions, the array configuration was not a critical factor in choosing when to perform observations.

The first set of observations were performed with the VLA in the D-array configuration, with the more southern sources observed in the intermediate DnC configuration, giving resolutions of about 15\arcsec\ (C-band) and 10\arcsec\ (X-band).  The widest available bandwith (50 MHz) was used.  Generally exposure times were around 2 minutes for C-band observations and around 3 minutes in X-band, although this varied depending on the brightness of the source at 1.4 GHz.  We aimed to obtain rms values of around 0.5 mJy or less for each source, and this was achieved except in a small number of cases where the rms was between 0.5 and 1 mJy.  Observations of a VLA flux calibrator were performed at least once an hour and phase calibrators located within at least 15\arcdeg\, and usually within 10\arcdeg\,of the science targets were done every half hour.  A little more than half of the combined BAL/normal quasar sample was observed in this configuration.

The second set of observations to complete the sample used the new EVLA in the B-array configuration, observing the southernmost sources in the hybrid BnA configuration.  The same frequencies were used, though the bandwidth on the EVLA is slightly larger at 64 MHz.  With this configuration the approximate resolutions are 1.5\arcsec\ (C-band) and 1\arcsec\ (X-band).  Again we aimed to achieve 0.5 mJy rms noise or better in the radio maps, and again except for a few rare cases this goal was reached.  The same calibrator strategy was used during this observing period.

Data at both frequencies were collected for all the sources in the sample, with a few exceptions: 3 sources (1 BAL quasar and 2 normal quasars) were only observed at one frequency either due to technical problems at the telescope or time constraints, and one BAL quasar was not observed at all for similar reasons, though it will be included in tables below for completeness.  Additionally, five BAL sources that overlap with the sample of Montenegro-Montes et al. (2008) were not observed by us, as discussed in the next section.

The data were calibrated using CASA\footnote{http://casa.nrao.edu/}, distributed by NRAO, using standard procedures one would use in AIPS for these straightforward continuum observations.  The flux density scale is that of Perley \& Taylor (1999). 

\subsection{Radio fluxes}
Radio maps for all sources were inspected and the 4.9 and 8.4 GHz integrated flux measurements of the BAL and normal quasar samples are listed in Tables~\ref{balpropstbl} (columns (8) and (9)) and~\ref{nbalpropstbl} (columns (7) and (8)).  In the case of resolved/extended sources, the integrated fluxes include the core and lobe emission.  Five BAL sources where the flux values from the Montenegro-Montes et al. (2008) sample were used are noted in the table.  Among their observations were simultaneous measurements at essentially the same frequencies as ours, so their inclusion as substitutes for our observations is valid.  Five more of their sources were also observed by us, and the values match well.  The few 4.9 and 8.4 GHz measurements that were not made simultaneously are also noted in this table.

In order to analyze our data in context with the rest of the radio spectrum, data at other frequencies were gathered from the literature where available.  In particular, we utilized fluxes from Montenegro-Montes et al (2008) (2.6, 4.8, 8.3, 10.5, and 15 GHz), the Green Bank 4.85 GHz Northern Sky Survey (GB6; Gregory et al. 1996), FIRST and NVSS (1.4 GHz), the Texas Survey of Radio Sources at 365 MHz (TEXAS; Douglas et al. 1996), the Westerbork Northern Sky Survey (WENSS, 325 MHz; Rengelink et al. 1997), and the VLA Low-Frequency Sky Survey (VLSS, 74 MHz; Cohen et al. 2007).  Plotted in Figures~\ref{balspectra} and~\ref{nbalspectra} are the radio spectra of the BAL and unabsorbed samples, respectively; our measurements are shown as asterisks, literature values overlapping (or close to) our observed frequencies are shown as open circles, and other values are shown as plusses.  Because the majority of our analysis utilizes the FIRST fluxes as opposed to those of NVSS, and for clarity in the spectra, only the FIRST values are plotted in these figures.

\subsection{Radio spectral indices}
The main goal of these observations was to measure the radio spectral index $\alpha$ of the two samples for comparison.  We did this in several ways, and would like to highlight the index measured between 4.9 and 8.4 GHz ($\alpha_{8.4}^{4.9}$) because the flux measurements are simultaneous and variability cannot affect the slope of the radio spectrum.  These values are given in column (10) of Table~\ref{balpropstbl} and column (9) of Table~\ref{nbalpropstbl}.  We also calculated the spectral index between 1.4 and 4.9 GHz ($\alpha_{4.9}^{1.4}$), and these are listed in column (11) of Table~\ref{balpropstbl} and column (10) of Table~\ref{nbalpropstbl}.  Finally, we did a simple linear fit to all available data points for each object (each source has at least 3 flux measurements), which is quite reasonable in most cases but, as is clear from inspecting the spectra in Figures~\ref{balspectra} and~\ref{nbalspectra}, some are more complex.  The spectral index measured in this way ($\alpha_{fit}$) is given in column (12) of Table~\ref{balpropstbl} and column (11) of Table~\ref{nbalpropstbl}.  Because the rms values in our radio maps and therefore the errors in the flux measurements are generally small, the errors on the spectral indices are also quite small and thus not reported in the table.

\section{RESULTS \& DISCUSSION}

\subsection{Radio spectra and spectral indices}
\begin{figure}
 \centering
  \figurenum{4}
   \includegraphics[width=6in]{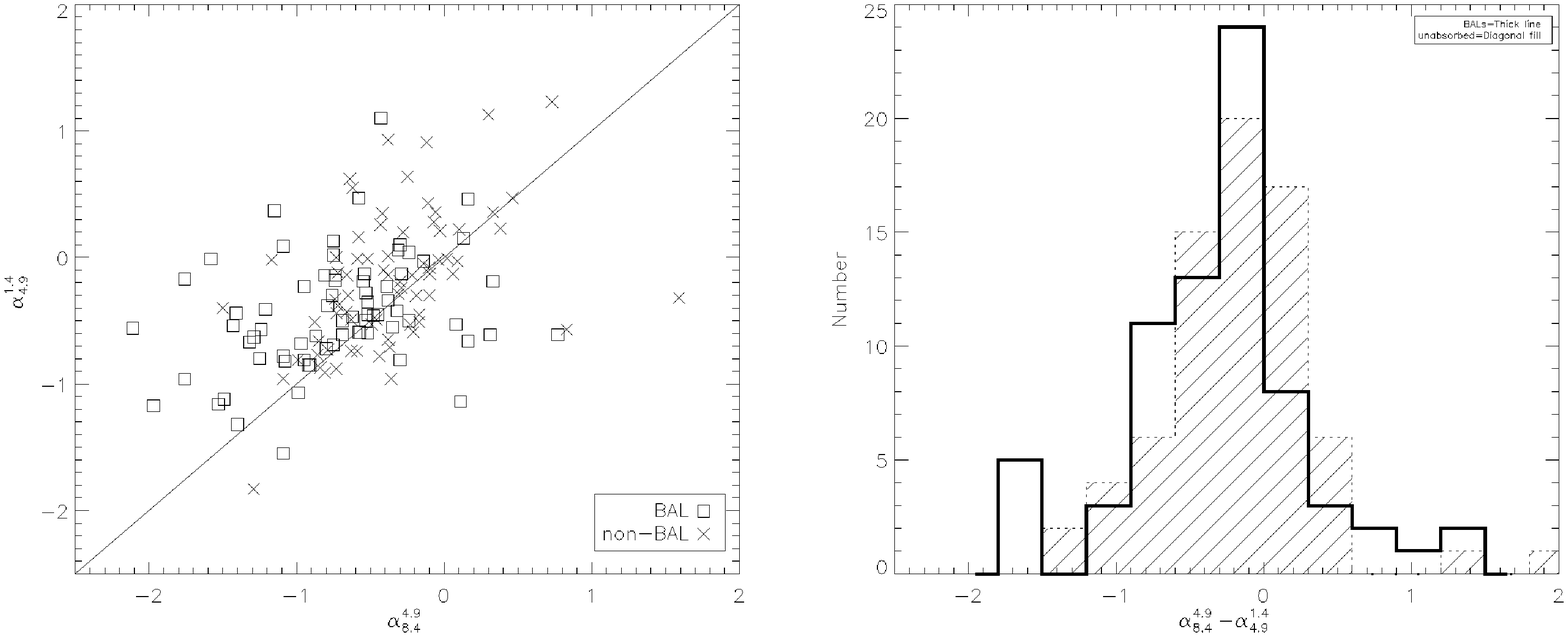}
  \caption{(Left) Comparison between $\alpha_{8.4}^{4.9}$ and $\alpha_{4.9}^{1.4}$.  The solid line indicates where the spectral index is the same in both regions of the spectrum.  In general, for both BAL and non-BAL quasars there is a flattening of the radio spectrum towards lower frequencies. (Right) Histograms of $\alpha_{8.4}^{4.9} - \alpha_{4.9}^{1.4}$ for both samples, illustrating that the flattening rate is similar in both samples.  A K-S test indicates that the distributions are from the same parent population.\label{colorcolorfig}}
\end{figure}

Because the majority of our sources only have data at three frequencies (1.4, 4.9, and 8.4 GHz), we have not attempted to locate peaks in the radio spectra or apply fits beyond the simple linear models (where all available data are included).  In the left panel of Figure~\ref{colorcolorfig} we plot a radio ``color-color'' plot, comparing the spectral indices $\alpha_{8.4}^{4.9}$ and $\alpha_{4.9}^{1.4}$.  The line is not a fit, just simply to illustrate where the two spectral indices would be equal.  We see that in general there tends to be a flattening of the radio spectrum toward lower frequencies in both samples, as the majority of points fall above the line.  It is possible that this is due to the presence of synchrotron self absorption in both samples, though again variability, especially at the higher 4.9 GHz frequency, cannot be completely ruled out.  This trend could also be an indication that BAL quasars do not show peaked spectra of the CSS/GPS type more often than normal quasars, but more data at lower frequencies for a larger number of objects is needed to do this properly.  To confirm that this flattening occurs at a similar rate in both samples, we also compared the distributions of the difference in spectral indices ($\alpha_{8.4}^{4.9}-\alpha_{4.9}^{1.4}$) for all objects in each sample (Figure~\ref{colorcolorfig}, right).  A statistical test on these distributions yields $P_{KS}=0.25$.  

One benefit of visualizing the data in this ``color-color'' plot is to identify strange or interesting sources.  We see one extreme outlier in this plot; the normal quasar SDSS J084617.52+375718.7 has an unusual highly positive slope at higher frequency, with $\alpha_{8.4}^{4.9}=1.59$.  While there are other objects with inverted slopes at these frequencies, none are nearly this steep.  We note that this is one of the few sources with non-simultaneous 4.9 and 8.4 GHz measurements, and so it is possible that it is variable at high frequency even though it does not appear to be variable at 1.4 GHz (see \S 4.3).  Another possibility is that there is a secondary component in this object contributing at high frequency only, though this would be unusual.  

We have also analyzed and compared the spectral indices  of the BAL and non-BAL samples as a group.  We remind the reader that the samples are one-to-one matched in redshift, and so the fact that we are comparing observed-frame spectral indices is not an issue.  The mean, median, and standard deviations of $\alpha_{8.4}^{4.9}$, $\alpha_{4.9}^{1.4}$, and $\alpha_{fit}$ are given in the top half of Table~\ref{alphastatstbl}.  The bottom half of the table gives these values when only considering compact ($\Theta < 0.1$; see \S 4.2) objects.  While the widths of the distributions are all quite similar, with $\sigma$ approximately between 0.4 and 0.5, it is clear that the means and medians of all measures of $\alpha$ for BAL quasars are lower (steeper) than those of non-BAL sources.  

\begin{figure}
 \centering
  \figurenum{5}
   \includegraphics[width=6in]{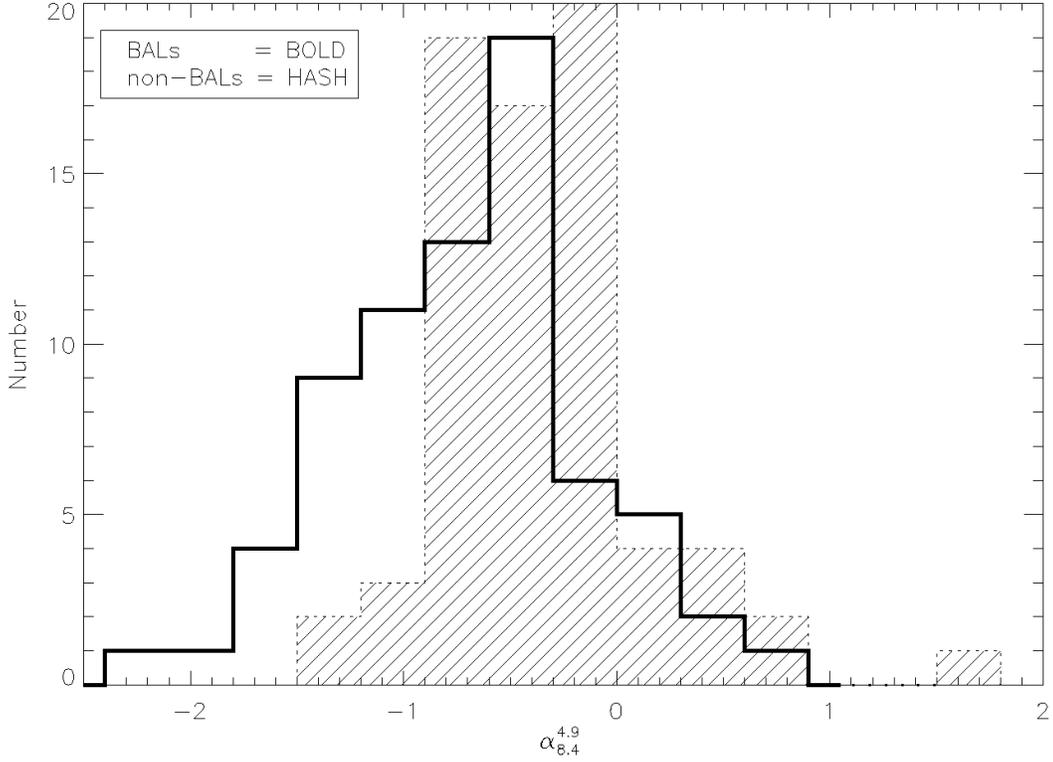}
  \caption{The radio spectral index $\alpha_{8.4}^{4.9}$ distributions for the BAL sample (bold) and non-BAL (diagonal fill) samples.  While both samples span a range of spectral indices, there is a significant favoring of steeper spectra for the BAL sample.  As a reminder to the reader, the flux measurements used here are simultaneous (with very few exceptions), and the redshifts of the samples are well matched so there would be very little difference if rest-frame spectral indices were used.  A K-S test gives $D_{ks}=0.347$, with a corresponding probability that the two distributions are from the same parent population of 0.0002.  A mean R-S test gives $Z_{rs}=4.00$, with a probability that the means are the same of $3.1 \times 10^{-5}$.\label{alphacxfig}}
\end{figure}

\begin{figure}
 \centering
  \figurenum{6}
   \includegraphics[width=6in]{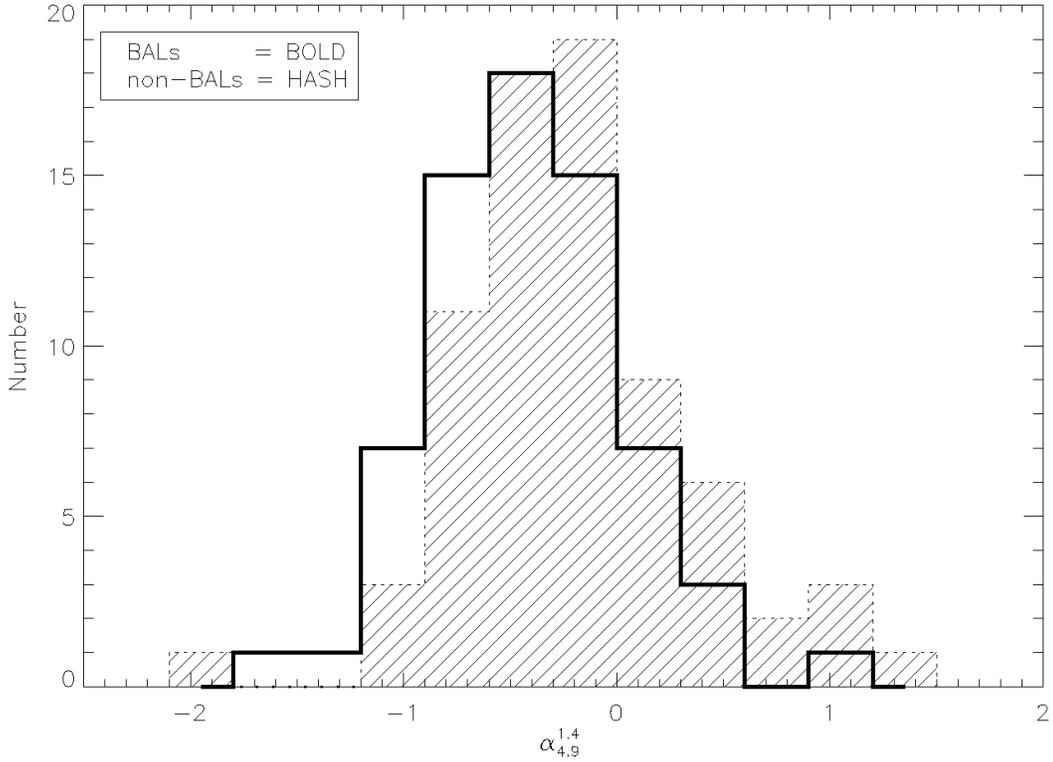}
  \caption{The radio spectral index $\alpha_{4.9}^{1.4}$ distributions for the BAL sample (bold) and non-BAL (diagonal fill) samples.  The favoring of steeper spectra for BAL quasars is not as significant but still present. A K-S test gives $D_{ks}=0.287$, with a corresponding probability that the two distributions are from the same parent population of 0.0036.  A mean R-S test gives $Z_{rs}=3.18$, with a probability that the means are the same of 0.0007. The flux measurements used to calculate $\alpha_{4.9}^{1.4}$ are not simultaneous, and so radio variability may affect the results.\label{alphaclfig}}
\end{figure}

\begin{figure}
 \centering
  \figurenum{7}
   \includegraphics[width=6in]{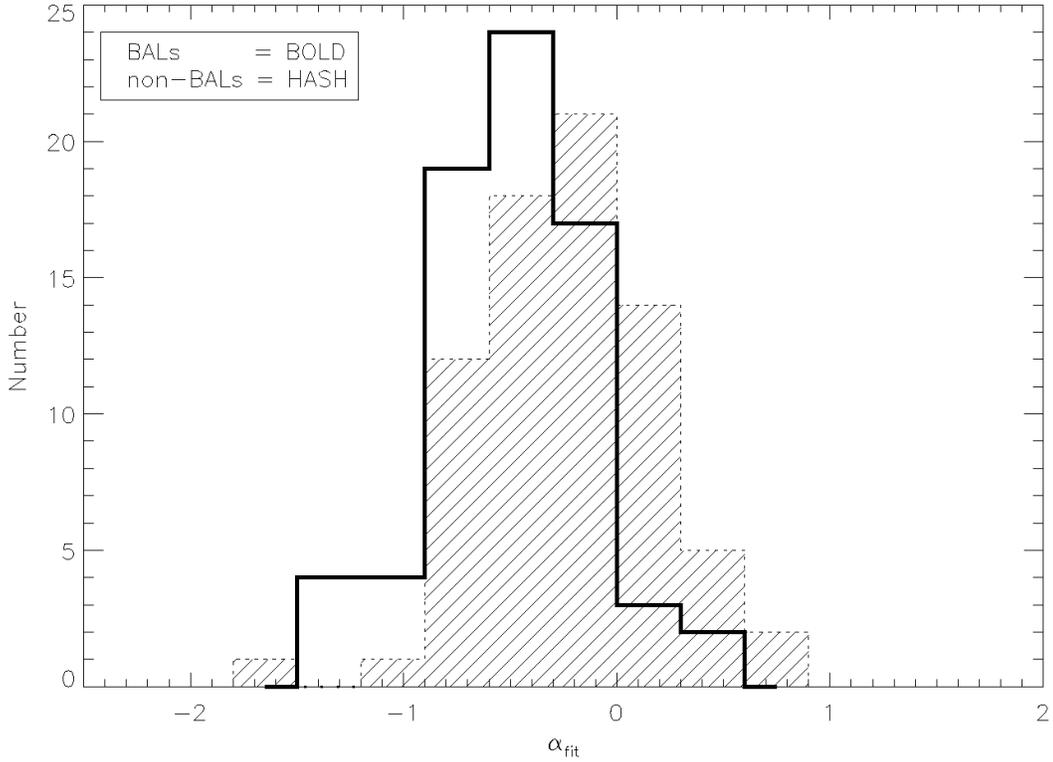}
  \caption{The radio spectral index $\alpha_{fit}$ distributions for the BAL sample (bold) and non-BAL (diagonal fill) samples.  The fits to the spectra are simple linear fits to all available data for each source.  The favoring of steeper spectra for BAL quasars is still present. A K-S test gives $D_{ks}=0.322$, with a corresponding probability that the two distributions are from the same parent population of 0.0007.  A mean R-S test gives $Z_{rs}=3.76$, with a probability that the means are the same of $8.4 \times 10^{-5}$.\label{alphafitfig}}
\end{figure}

Figures~\ref{alphacxfig}, ~\ref{alphaclfig}, and~\ref{alphafitfig} graphically show the distributions of $\alpha_{8.4}^{4.9}$, $\alpha_{4.9}^{1.4}$, and $\alpha_{fit}$, respectively, for the BAL and non-BAL samples.  Just by visual inspection of the distributions of $\alpha_{8.4}^{4.9}$ we can see that there is a favoring of steeper spectra in the BAL quasars, or at least an overabundance of BALs with $\alpha_{8.4}^{4.9}$ between $-1$ and $-2$, though as mentioned previously both samples do have a wide range.  The odd non-BAL with a highly inverted high frequency spectrum that was discussed above is also a clear outlier in this plot.  The shift between the two samples is apparent, though not as strong, when examining the distributions of the spectral index $\alpha_{4.9}^{1.4}$.  As noted variability issues cannot be ruled out in this comparison, at 4.9 GHz in particular, and may account for the difference not appearing as striking in this comparison.  Regardless, what we wish to highlight is that the distributions are measurably different, as shown below.  We do acknowledge that using only two closely spaced points in the spectrum for this analysis can be problematic.  However, the difference in the distributions using all available data ($\alpha_{fit}$) is also clearly apparent, with BAL quasars favoring steeper spectra.

Both Kolmogorov-Smirnov (K-S) tests  and Wilcoxon Rank-Sum (R-S) tests have been performed on the three spectral index measurements made, and the results are shown in Table~\ref{stattesttbl}.  $P_{ks}$ is the probability that the two samples are from the same parent population from the K-S test, and $P_{rs}$ is the probability that the distribution means are the same from the R-S test.  In all cases and by both tests the differences are significant, though most strongly in the cases of $\alpha_{8.4}^{4.9}$ and $\alpha_{fit}$.  When restricted to compact sources (these are the values in the bottom half of the table) the significance remains, though less strongly for $\alpha_{8.4}^{4.9}$.  It is possible that this reduction in significance is simply due to the decreased sample size.  It is worth noting that when this comparison was done after the first period of observations, the difference was apparent but uncertain.  As more data came in from the second period, the difference grew stronger, suggesting that the scatter in the relationship between $\alpha$ and orientation is large enough that large samples such as this one are required to make meaningful comparisons.  Finally, since our sample consists of matched pairs of BAL and non-BAL quasars, we also performed a signed rank-sum test on the difference between each measure of spectral index for each pair (matched-pair test).  The resulting probabilities that the difference distributions are symmetric about 0 (and therefore that the samples are indistinguishable) are $3.7 \times 10^{-5}$ ($\alpha_{8.4}^{4.9}$), 0.0022 ($\alpha_{4.9}^{1.4}$), and 0.0005 ($\alpha_{fit})$.  While we have mentioned issues regarding variability, complexities in the spectral shapes, and the spacing between some of the measurements, the fact that all statistical tests on all three measurements of $\alpha$ show a significant difference, even when only considering compact sources, strongly suggests that the difference between the two samples is real.

We have also looked at the spectral index distributions of HiBAL and LoBAL subsamples, and there is no clear difference between the two.  However, the numbers are too small to be conclusive, as there are only 11 Lo/FeLoBALs in the sample.

Knowing that spectral index is an orientation indicator, at least in a statistical sense for a sample, these results indicate that while BAL quasars can have a range of orientations (as suggested by the widths of the distribution), they seem to more often be seen ``edge-on'' and therefore exhibit somewhat steeper radio spectra.  Normal, unabsorbed quasars, then are generally seen more ``face-on'', though again they likely span a range of orientations.  This result illustrates the complexity of the problem- it is clear that the edge-on only, simple orientation schemes cannot fully explain these objects, but orientation apparently still plays an important role in the presence or absence of BAL features.  This is an important result that has not yet been directly observed, though some modeling has suggested it (Shankar et al. 2008); therefore, we will highlight it, provide a more quantitative analysis of its implications, and engage in a more detailed discussion in a forthcoming paper (DiPompeo et al., in preparation).

\subsection{Morphology}
It has been noted that BAL quasars are generally more compact in radio maps than normal quasars, as discussed in \S 1.  Using the ``morphological parameter'' $\Theta$, defined by Montenegro-Montes et al. (2008) as $\Theta = \log (S_{i}/S_{p})$, for the FIRST fluxes in our sample, we do not see the significant difference seen elsewhere.  Assuming $\Theta > 0.1$ is considered extended, then 10 BAL quasars (13.5\%) and 16 (21.6\%) normal quasars have structure resolved by FIRST.  The number of extended BAL quasars is consistent with previous results at similar resolution (for example, about 10\% in Becker et al. 2000), but the number of extended non-BAL quasars is lower than expected.  This could be because our selection of unabsorbed quasars when building the matched sample may be biased against those with individually resolved components, since they are solely based on a match in the FIRST catalog with an integrated flux within 20\% of the corresponding BAL quasar, which will not include flux from individual components.  Therefore the only extended non-BAL sources that will make it into the sample are those with cores that have a similar flux to their respective BAL quasar, likely under-representing extended objects in the sample.  So while it is clear here that normal quasars do more often exhibit extended structure, we cannot be certain to what extent.

For the subset of our observations that had higher resolution than FIRST, we found no evidence of newly resolved structure and therefore have no new morphological information to present.

\subsection{Variability \& Polar BALs?}
\begin{figure}
 \centering
  \figurenum{8}
   \includegraphics[width=6in]{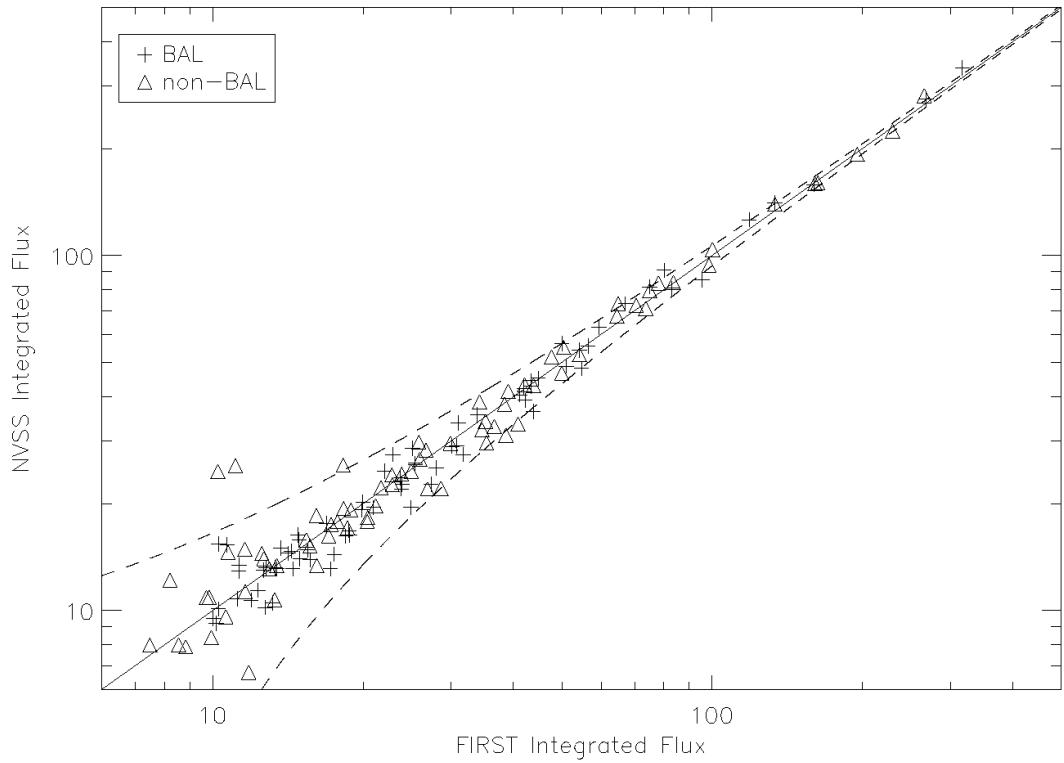}
  \caption{Comparing the FIRST and NVSS fluxes as a check for variability.  The solid line shows where the two survey fluxes are equal, and the dotted lines show the 3-$\sigma$ variation of the sources around the line of equal flux.  The two non-BAL sources that seem to lie well above the line may both have a second object contaminating the NVSS fluxes, and are probably not actually variable.\label{variablefig}}
\end{figure}

A check for radio variability was done by comparing the FIRST and NVSS fluxes, since the surveys were performed at different times.  As some of our tests combine data taken at different times, the extent and role of variability may be important.  Figure~\ref{variablefig} shows the comparison between FIRST and NVSS fluxes; the solid line indicates where the two survey fluxes are equal, and the dotted lines show the 3-$\sigma$ variation of the points around the line of equal flux.  From this comparison we see that most sources fall close to this line and thus do not appear to be significantly variable.  The two non-BAL quasars (SDSS J083629.57+345544.8 and SDSS J105611.77+315616.5) that seem to have much larger NVSS fluxes and thus lie well above the line may have contaminating sources about 35\arcsec\ away, and so the differences in flux are likely due to resolution issues and not real variability.  At the frequencies we observed, the resolution is sufficient that contamination of these nearby sources is not an issue.

For a quantitative check, we also compute the variability parameter as defined by Torniainen et al. (2005):
$$Var_{\Delta S} = \frac{S_{max}-S_{min}}{S_{min}}.$$
The significance of the variation was computed using a modified method of Zhou et al. (2006) in which the integrated fluxes from both surveys are used, instead of the peak flux from FIRST (as the variable sources are likely to be seen more pole-on and thus appear point-like, this should make little difference):
$$\sigma_{var} = \frac{|S_2-S_1|}{\sqrt{\sigma_2^2+\sigma_1^2}}$$
where $\sigma_1$ and $\sigma_2$ are the uncertainties in the integrated fluxes.  The error in integrated flux from FIRST here is assumed to be 5\%, since only peak rms values are reported in the survey.  We chose a conservative value of $\sigma_{var} > 4$ to suggest real variability.  Sources satisfying this criteria (2 BAL quasars and 4 unabsorbed quasars) are listed in Table~\ref{vartbl}.  We however note that at higher frequency variability may be more important.  We do not identify any of the sources from Ghosh \& Punsly (2007) or Zhou et al. (2006) as significantly variable, though there is overlap between our samples, due both to the use here of integrated fluxes from both surveys as well as the more strict cut of $\sigma_{var} > 4$.

Two BAL sources have evidence of variability on short timescales, a property that has been used to suggest viewing angles within 20\arcdeg\ of the jet axis in order to avoid the inverse Compton catastrophe.  While using FIRST and NVSS for this sort of analysis is useful because of the large number of sources and overlap between the two, we are cautious of this method because it may be extracting more information from these surveys than they were meant to provide- a few mJy difference between sources could just be noise as the absolute calibrations are simply not that accurate.  Both of the BAL quasars identified here only differ in flux between the surveys by about 5 mJy.  However, for completeness we also calculated the brightness temperature $T_b$ of the two possibly variable BAL quasars in our sample using equation 4 of Ghosh \& Punsly (2007): 125243.85+005320.1 has $T_b = 5.7 \times 10^{12}$ K and 135550.30+361627.6  has $T_b = 2.3 \times 10^{14}$ K.  Both of these are greater than the $10^{12}$ K above which the inverse Compton catastrophe should occur, and suggests that these could also be polar BAL quasars.  It should also be noted that the date of observation for an object in FIRST (an important parameter in $T_b$) is difficult to know with certainty, as the fields overlap and the final fluxes presented are a combination of data that could have been taken days, weeks, or even years apart.  This is not likely to be a large factor in the case of sources with $T_b$ much larger than $10^{12}$ K, but for borderline objects (as one of ours is), it could play an important  role.

The numbers are small, but it is interesting to note that of the ``polar'' BALs identified in the literature that are in our sample, as well as the two just discussed (for 8 total), only two of them have what would technically be considered flat radio spectra (using $\alpha_{8.4}^{4.9} > -0.5$).  We would of course expect that if these are truly face-on objects that they would show a preference for flat spectra.  This may be support for the idea that in many cases the variability used to identify polar BALs is due more to survey limitations than real quasar properties, but would need to be examined in larger numbers to be more conclusive.

\section{Summary}
We have observed a large number of BAL quasars (74) and a well matched sample of 74 unabsorbed quasars at 4.9 and 8.4 GHz, simultaneously in all but a few cases.  All sources also have data at 1.4 GHz available from the FIRST survey, and other data from 74 MHz to 15 GHz have been collected from the literature when available.  The main results are as follows:

\begin{enumerate}
\item About 13\% of the BAL sample appears extended in FIRST maps, similar to previous findings at arcsecond resolutions.  Only 21\% of the non-BAL sample show resolved structure, lower than previously found- however, there is likely a bias in our selection of unabsorbed quasars against those with individually resolved components.  

\item We identify two new BAL quasars that may be significantly variable at 1.4 GHz with brightness temperatures both above $10^{12}$ K, suggesting they may be viewed pole-on.  

\item Analysis of the spectral index between 4.9 and 8.4 GHz (simultaneous), 1.4 and 4.9 GHz (non-simultaneous), and using a simple linear fit to all available data shows that the two samples do significantly differ, with the level of significance depending on the frequencies analyzed and statistical test performed, but always above 3-$\sigma$.  BAL quasars do in fact show an overabundance of steep spectrum sources compared to non-BAL quasars, though both samples span a wide range of spectral index.  This difference persists when only compact objects are considered.  It may be that this difference has not been seen before because the sample sizes were simply too small.  The simplest interpretation is that although BAL quasars may span a range of orientations, they do have a preference for larger viewing angles.  Clearly the simplest orientation models have been ruled out, but that is not to say that orientation does not play a role in these objects.

\end{enumerate}

\acknowledgments
M. DiPompeo would like to thank the European Southern Observatory for providing DGDF funding to support a visit to collaborate with C. De Breuck in 2011, as well as the excellent support staff at NRAO that was extremely helpful during the calibration process using CASA.

\begin{deluxetable}{rcrrccccccccc}
 \tabletypesize{\scriptsize}
 \tablewidth{0pt}
 \tablecaption{The BAL quasar sample\label{balpropstbl}}
 \tablehead{
   \colhead{Source Name} & \colhead{Type} & \colhead{RA} & \colhead{DEC} & \colhead{$z$} & \colhead{$M_B$} & \colhead{$S_{1.4}$} & \colhead{$S_{4.9}$} & \colhead{$S_{8.4}$} & \colhead{$\alpha_{8.4}^{4.9}$} & 		      \colhead{$\alpha_{4.9}^{1.4}$} & \colhead{$\alpha_{fit}$}   \\
     \colhead{(SDSS J)} &  & \colhead{} &\colhead{} &  \colhead{} &  \colhead{} & \colhead{(mJy)} & \colhead{(mJy)}  &  \colhead{(mJy)} & \colhead{}  & \colhead{}  & \colhead{}  \\
   \colhead{(1)} &  \colhead{(2)} &  \colhead{(3)} &  \colhead{(4)} &  \colhead{(5)} &  \colhead{(6)} &  \colhead{(7)} &  \colhead{(8)} &  \colhead{(9)}  &  \colhead{(10)}  &  \colhead{(11)}  &  \colhead{(12)}
  }
   \startdata
 001408.22$-$085242.2 & Hi & 00 14 08.2 &$-$08 52 42.3 & 1.74 &$-$25.8 & 14.3 & 17.4                  & 18.7                  &  0.13    &  0.15   &  0.15   \\
 002440.99+004557.7   & Hi & 00 24 40.9 &  00 45 57.1 & 2.24 &$-$25.7 & 59.3 & 21.4                  & 12.8                  &$-$0.95    &$-$0.81   &$-$0.84   \\
 003923.18$-$001452.6 & Hi & 00 39 23.1 &$-$00 14 52.6 & 2.23 &$-$25.9 & 20.9 & 17.6                  & 11.4                  &$-$0.81    &$-$0.14   &$-$0.29   \\
 004444.06+001303.5   & Hi & 00 44 44.0 &  00 13 03.5 & 2.29 &$-$25.7 & 54.7 & 30.8                  & 23.8                  &$-$0.48    &$-$0.46   &$-$0.46   \\
 014847.61$-$081936.3 & Lo & 01 48 47.6 &$-$08 19 36.3 & 1.68 &$-$26.2 & 74.9 & 17.6                  &  7.7                  &$-$1.53    &$-$1.16   &$-$1.21   \\
 024534.07+010813.7   & Hi & 02 45 34.0 &  01 08 13.7 & 1.54 &$-$25.8 &317.1 & 78.2                  & 35.1                  &$-$1.49    &$-$1.12   &$-$1.03   \\
 075310.42+210244.3   & Lo & 07 53 10.4 &  21 02 44.3 & 2.29 &$-$26.4 & 17.5 &  8.6                  &  4.4                  &$-$1.24    &$-$0.57   &$-$0.73   \\
 080351.63+500317.6   & Hi & 08 03 51.6 &  50 03 17.7 & 2.95 &$-$26.4 & 13.4 &  3.2                  &  3.4                  &  0.11    &$-$1.14   &$-$0.79   \\
 081102.91+500724.5   & Hi & 08 11 02.9 &  50 07 24.5 & 1.84 &$-$26.6 & 24.9 & 10.9                  & 11.9                  &  0.16    &$-$0.66   &$-$0.41   \\
 082813.47+065326.3   & Hi & 08 28 13.4 &  06 53 26.2 & 3.00 &$-$27.3 & 19.8 & 10.6                  &  9.3                  &$-$0.24    &$-$0.50   &$-$0.47   \\
 083749.59+364145.4   & Lo & 08 37 49.5 &  36 41 45.6 & 3.42 &$-$27.0 & 27.1 & 20.2                  & 12.1                  &$-$0.95    &$-$0.23   &$-$0.44   \\
 084158.47+392120.9   & Hi & 08 41 58.6 &  39 21 14.7 & 2.04 &$-$26.3 & 14.8 & 8.83                  &  4.6                  &$-$1.21    &$-$0.41   &$-$0.55   \\
 084224.38+063116.7   & Hi & 08 42 24.3 &  06 31 16.7 & 2.46 &$-$26.2 & 51.0 & 25.6                  & 21.2                  &$-$0.35    &$-$0.55   &$-$0.49   \\
 084303.93+335833.3   & Hi & 08 43 03.9 &  33 58 33.2 & 1.83 &$-$26.1 & 10.2 &  8.2                  &  5.5                  &$-$0.74    &$-$0.18   &$-$0.18   \\
 084914.27+275729.7   & Hi & 08 49 14.2 &  27 57 29.9 & 1.73 &$-$26.4 & 67.0 & 28.5                  & 16.9                  &$-$0.97    &$-$0.68   &$-$0.74   \\
 085417.61+532735.3   & Hi & 08 54 17.6 &  53 27 35.1 & 2.42 &$-$29.1 & 22.0 & 14.3                  & 10.8                  &$-$0.52    &$-$0.35   &$-$0.41   \\
 085641.58+424254.1   & Hi & 08 56 41.5 &  42 42 53.9 & 3.06 &$-$28.1 & 19.9 & 20.5                  & 13.7                  &$-$0.75    &  0.02   &$-$0.19   \\
 090552.40+025931.4   & Hi & 09 05 52.4 &  02 59 31.6 & 1.82 &$-$28.5 & 43.8 & 34.7                  & 25.8                  &$-$0.55    &$-$0.19   &$-$0.28   \\
 092556.56+063015.8   & Hi & 09 25 56.5 &  06 30 15.7 & 2.47 &$-$25.9 & 16.9 & 16.3                  & 15.1                  &$-$0.14    &$-$0.03   &$-$0.06   \\
 092913.96+375742.9   & Hi & 09 29 13.9 &  37 57 42.9 & 1.92 &$-$27.7 & 43.4 & 28.3                  & 23.0                  &$-$0.38    &$-$0.34   &$-$0.35   \\
 093348.37+313335.2   & Lo & 09 33 48.3 &  31 33 35.3 & 2.60 &$-$26.8 & 18.8 & 10.0                  &  7.5                  &$-$0.53    &$-$0.50   &$-$0.51   \\
 093559.50+525202.0   & Hi & 09 35 59.5 &  52 52 02.0 & 2.07 &$-$25.6 & 14.4 &  7.7                  &  5.3                  &$-$0.69    &$-$0.50   &$-$0.54   \\
 093804.52+120011.4   & Hi & 09 38 04.5 &  12 00 11.2 & 2.23 &$-$27.3 & 12.7 & 14.2\tablenotemark{a} &  7.9\tablenotemark{a} &$-$1.09    &  0.09   &$-$0.21   \\
 094732.09+325220.5   & Hi & 09 47 32.0 &  32 52 20.6 & 2.34 &$-$27.1 & 23.8 & 13.7                  &  6.4                  &$-$1.41    &$-$0.44   &$-$0.66   \\
 095327.96+322551.6   & Hi & 09 53 27.9 &  32 25 51.6 & 1.57 &$-$27.8 &133.4 &  237.0              &  259.0             &  0.16    &  0.46   &  0.39   \\
 095929.88+633359.8   & Hi & 09 59 29.8 &  63 33 59.8 & 1.85 &$-$28.5 & 15.5 & 13.1                  &  9.8                  &$-$0.54    &$-$0.13   &$-$0.25   \\
 100109.41+114608.8   & Lo & 10 01 09.4 &  11 46 08.7 & 2.28 &$-$26.9 & 23.7 & 16.8                  & 12.6                  &$-$0.53    &$-$0.28   &$-$0.35   \\
 102258.41+123429.7   & Hi & 10 22 58.4 &  12 34 26.2 & 1.73 &$-$26.8 &118.6 & 44.5                  & 24.7                  &$-$1.09    &$-$0.78   &$-$0.81   \\
 103351.87$-$002414.5 & Hi & 10 33 52.2 &$-$00 24 18.2 & 1.65 &$-$26.4 & 31.0 & 11.1                  &  6.2                  &$-$1.08    &$-$0.82   &$-$0.87   \\
 104059.79+055524.4   & Hi & 10 40 59.8 &  05 55 24.7 & 2.44 &$-$26.6 & 42.2 &  8.1                  &  3.8                  &$-$1.40    &$-$1.32   &$-$1.34   \\
 104452.41+104005.9   & Hi & 10 44 52.4 &  10 40 05.9 & 1.88 &$-$27.8 & 17.2 &  5.9                  &  3.6                  &$-$0.92    &$-$0.85   &$-$0.86   \\
 105352.86$-$005852.7 & Lo & 10 53 52.8 &$-$00 58 52.5 & 1.50 &$-$26.9 & 27.3 & 15.7\tablenotemark{c} & 11.8\tablenotemark{c} &$-$0.52    &$-$0.45   &$-$0.47   \\
 105416.51+512326.0   & Hi & 10 54 16.5 &  51 23 26.2 & 2.34 &$-$27.7 & 33.8 & 14.2                  &  9.5                  &$-$0.75    &$-$0.69   &$-$0.70   \\
 110206.66+112104.9   & Hi & 11 02 06.6 &  11 21 04.7 & 2.35 &$-$28.0 & 83.0 & 35.8                  & 17.6                  &$-$1.32    &$-$0.67   &$-$0.78   \\
 110531.41+151215.9   & Hi & 11 05 31.4 &  15 12 17.5 & 2.07 &$-$27.3 & 12.3 &  4.5                  &  2.3                  &$-$1.25    &$-$0.80   &$-$0.88   \\
 112241.46+303534.9   & Lo & 11 22 41.4 &  30 35 34.8 & 1.81 &$-$28.6 & 10.0 & 10.6                  &  9.3                  &$-$0.24    &  0.04   &$-$0.04   \\
 112938.47+440325.0   & Hi & 11 29 38.4 &  44 03 25.0 & 2.21 &$-$28.2 & 42.0 & 35.1                  & 23.5                  &$-$0.74    &$-$0.14   &$-$0.27   \\
 113152.56+584510.2   & Hi & 11 31 52.5 &  58 45 10.2 & 2.26 &$-$27.1 & 12.7 &  6.5\tablenotemark{a} &  3.0\tablenotemark{a} &$-$1.43    &$-$0.54   &$-$0.70   \\
 113445.83+431858.0   & Lo & 11 34 45.8 &  43 18 57.8 & 2.18 &$-$25.3 & 28.0 & 13.4                  &  9.8                  &$-$0.58    &$-$0.59   &$-$0.58   \\
 115901.75+065619.0   & Hi & 11 59 01.7 &  06 56 18.9 & 2.19 &$-$26.5 &160.1 & 55.2                  & 33.8                  &$-$0.91    &$-$0.85   &$-$0.86   \\
 115944.82+011206.9   & Hi & 11 59 44.8 &  01 12 06.8 & 2.00 &$-$28.5 &268.4 &  125.0               &  147.0              &  0.31    &$-$0.61   &$-$0.35   \\
 121323.94+010414.7   & Hi & 12 13 23.9 &  01 04 14.8 & 2.83 &$-$26.2 & 22.9 & 15.8                  & 10.5                  &$-$0.76    &$-$0.30   &$-$0.39   \\
 122848.21$-$010414.5 & Hi & 12 28 48.1 &$-$01 04 14.2 & 2.66 &$-$28.5 & 30.8 & 17.6                  & 13.8                  &$-$0.45    &$-$0.45   &$-$0.44   \\
 123411.73+615832.6   & Hi & 12 34 11.7 &  61 58 32.4 & 1.95 &$-$26.9 & 23.9 & 14.9                  &  9.7                  &$-$0.79    &$-$0.38   &$-$0.50   \\
 123511.59+073330.7   & Hi & 12 35 11.6 &  07 33 30.8 & 3.03 &$-$27.9 & 11.2 &  2.6                  &  0.9                  &$-$1.97    &$-$1.17   &$-$1.25   \\
 123548.54+415659.4   & Hi & 12 35 48.5 &  41 56 59.4 & 1.52 &$-$25.7 & 15.6 &  4.1                  &  2.4                  &$-$0.99    &$-$1.07   &$-$1.03   \\
 123717.44+470807.0   & Hi & 12 37 17.4 &  47 08 07.0 & 2.27 &$-$27.3 & 80.1 & 63.2                  & 75.7                  &  0.33    &$-$0.19   &$-$0.10   \\
 123954.15+373954.5   & Hi & 12 39 54.1 &  37 39 54.4 & 1.84 &$-$25.4 & 18.4 &  8.4\tablenotemark{a} &  4.2\tablenotemark{a} &$-$1.29    &$-$0.63   &$-$0.71   \\
 125243.85+005320.1   & Hi & 12 52 43.8 &  00 53 20.2 & 1.69 &$-$27.3 & 10.2 & 10.1                  &  4.3                  &$-$1.58    &$-$0.01   &$-$0.45   \\
 130208.27$-$003731.6 & Lo & 13 02 08.2 &$-$00 37 31.3 & 1.67 &$-$26.8 & 11.9 & 19.0\tablenotemark{b} & 10.2\tablenotemark{b} &$-$1.15    &  0.37   &$-$0.10   \\
 130448.06+130416.5   & Hi & 13 04 48.0 &  13 04 16.6 & 2.57 &$-$26.5 & 50.0 & 23.2                  & 16.0                  &$-$0.69    &$-$0.61   &$-$0.66   \\
 130756.73+042215.5   & Hi & 13 07 56.7 &  04 22 15.5 & 3.02 &$-$28.6 & 14.9 & 11.7                  &  8.7                  &$-$0.55    &$-$0.19   &$-$0.27   \\
 133701.39$-$024630.3 & Hi & 13 37 01.3 &$-$02 46 29.8 & 3.06 &$-$27.8 & 44.8 & 36.0                  & 13.9                  &$-$1.76    &$-$0.17   &$-$0.63   \\
 135550.30+361627.6   & Hi & 13 55 50.2 &  36 16 27.5 & 1.86 &$-$25.8 & 10.6 & 19.3                  & 14.1                  &$-$0.58    &  0.47   &  0.29   \\
 135910.77+400218.6   & Hi & 13 59 10.7 &  40 02 18.6 & 2.01 &$-$27.2 & 14.9 & 16.1                  & 13.6                  &$-$0.31    &  0.06   &  0.01   \\
 140736.65+483737.5   & Hi & 14 07 36.6 &  48 37 37.5 & 1.59 &$-$26.0 & 30.1 & 13.9                  &  8.7                  &$-$0.87    &$-$0.62   &$-$0.68   \\
 141334.38+421201.7   & Hi & 14 13 34.4 &  42 12 01.7 & 2.82 &$-$27.9 & 18.7 &  8.8\tablenotemark{c} & 13.4\tablenotemark{c} &  0.77    &$-$0.61   &$-$0.18   \\
 141445.72+544425.6   & Hi & 14 14 45.7 &  54 44 25.6 & 1.54 &$-$26.7 & 13.3 & 15.6                  & 10.4                  &$-$0.75    &  0.13   &$-$0.13   \\
 143340.35+512019.3   & Lo & 14 33 40.3 &  51 20 19.7 & 1.86 &$-$26.5 & 12.6 &  1.8                  &  1.0                  &$-$1.09    &$-$1.55   &$-$1.39   \\
 144434.80+003305.3   & Hi & 14 44 34.8 &  00 33 05.4 & 2.04 &$-$26.6 & 13.1 &  6.8                  &  7.1                  &  0.08    &$-$0.53   &$-$0.35   \\
 144707.40+520340.1   & Hi & 14 47 09.4 &  52 03 27.1 & 2.06 &$-$28.1 & 43.3 & 13.1                  & \nodata               & \nodata  &$-$0.95   &$-$0.99   \\
 145910.13+425213.2   & Hi & 14 59 10.1 &  42 52 13.2 & 2.97 &$-$28.5 & 13.6 & 15.5                  & 13.2                  &$-$0.30    &  0.10   &$-$0.02   \\
 150332.93+440120.6   & Hi & 15 03 32.9 &  44 01 20.6 & 2.05 &$-$26.8 & 11.2 &  8.4                  &  6.8                  &$-$0.39    &$-$0.23   &$-$0.27   \\
 151630.30$-$005625.5 & Hi & 15 16 30.3 &$-$00 56 24.6 & 1.92 &$-$26.7 & 25.4 & 14.1                  & 10.1                  &$-$0.62    &$-$0.47   &$-$0.51   \\
 153729.54+583224.6   & Hi & 15 37 29.5 &  58 32 24.7 & 3.06 &$-$25.6 & 14.1 & 56.3                  & 44.7                  &$-$0.43    &  1.10   &  0.56   \\
 154439.70+444050.9   & Hi & 15 44 39.6 &  44 40 50.4 & 1.56 &$-$26.0 & 25.1 &  7.5                  &  2.9                  &$-$1.76    &$-$0.96   &$-$1.00   \\
 155429.40+300118.9   & Hi & 15 54 29.4 &  30 01 19.0 & 2.69 &$-$28.7 & 41.2 & 34.9                  & 29.8                  &$-$0.29    &$-$0.13   &$-$0.18   \\
 155633.77+351757.3   & Fe & 15 56 33.7 &  35 17 57.6 & 1.05 &$-$24.0 & 31.7 & 18.7                  & 15.7                  &$-$0.32    &$-$0.42   &$-$0.39   \\
 160354.15+300208.6   & Hi & 16 03 54.1 &  30 02 08.8 & 2.03 &$-$28.0 & 54.1 & 26.1\tablenotemark{c} & 19.1\tablenotemark{c} &$-$0.57    &$-$0.59   &$-$0.56   \\
 162453.47+375806.6   & Hi & 16 24 53.4 &  37 58 06.6 & 3.38 &$-$28.0 & 56.4 & 23.3\tablenotemark{c} & 15.0\tablenotemark{c} &$-$0.80    &$-$0.72   &$-$0.59   \\
 162559.90+485817.5   & Hi & 16 25 59.9 &  48 58 17.5 & 2.72 &$-$28.6 & 25.4 &  9.4\tablenotemark{c} &  8.0\tablenotemark{c} &$-$0.30    &$-$0.81   &$-$0.67   \\
 162656.74+295328.0   & Hi & 16 26 56.7 &  29 53 28.0 & 2.31 &$-$27.2 & 11.3 &  5.6                  &  1.8                  &$-$2.11    &$-$0.56   &$-$0.80   \\
 165543.24+394519.9   & Hi & 16 55 43.2 &  39 45 19.9 & 1.75 &$-$27.4 & 10.1 & \nodata               & \nodata               & \nodata  & \nodata & \nodata \\
 170558.99+243532.6   & Hi & 17 05 59.0 &  24 35 32.7 & 1.56 &$-$25.3 & 95.4 & 44.7                  & 33.7                  &$-$0.52    &$-$0.60   &$-$0.59   \\
    \enddata
 \tablecomments{Properties of the of the BAL sample. Column (1) is the SDSS name of the source, (2) is it's BAL subtype, (3) and (4) are the J2000 RA and DEC (radio positions from FIRST), (5) is the SDSS redshift, (6) is the absolute B magnitude, (7) is the FIRST 1.4 GHz integrated flux, (8) and (9) are our new 4.9 and 8.4 GHz measurements.  Columns (10) and (11) are two-point spectral indices between 4.9 and 8.4 GHz and 1.4 and 4.9 GHz, respectively.  Column (11) is the spectral index using a linear fit to our new measurements combined with all other data available in the literature.}
 \tablenotetext{a}{Observations not simultaneous- 8.4 GHz observed with VLA in 2010, 4.9 GHz observed with EVLA in 2011.}
 \tablenotetext{b}{Observations not simultaneous- 4.9 GHz observed with VLA in 2010, 8.4 GHz observed with EVLA in 2011.}
 \tablenotetext{c}{Values from Montenegro-Montes et al. (2008) at 4.8 GHz and 8.3 GHz.  These observations were made essentially simultaneously.}
\end{deluxetable}

\clearpage

\begin{deluxetable}{rrrcccccccc}
 \tabletypesize{\scriptsize}
 \tablewidth{0pt}
 \tablecaption{The unabsorbed quasar sample\label{nbalpropstbl}}
 \tablehead{
   \colhead{Source Name} & \colhead{RA} & \colhead{DEC} & \colhead{$z$} & \colhead{$M_B$} & \colhead{$S_{1.4}$} & \colhead{$S_{4.9}$} & \colhead{$S_{8.4}$} & \colhead{$\alpha_{8.4}^{4.9}$} & \colhead{$\alpha_{4.9}^{1.4}$} & \colhead{$\alpha_{fit}$}  \\
    \colhead{(SDSS J)} &  \colhead{} &  \colhead{} &  \colhead{} &  \colhead{} &  \colhead{(mJy)} &  \colhead{(mJy)} &  \colhead{(mJy)}  & \colhead{}  & \colhead{}  & \colhead{}   \\
    \colhead{(1)} &  \colhead{(2)} &  \colhead{(3)} &  \colhead{(4)} &  \colhead{(5)} &  \colhead{(6)} &  \colhead{(7)} &  \colhead{(8)}   &  \colhead{(9)} &  \colhead{(10)} &  \colhead{(11)}
  }
   \startdata
 000050.60$-$102155.9 & 00 00 50.6 &$-$10 21 56.0 & 2.64 &$-$28.5 & 20.3 & 12.7                  &  8.6                  &$-$0.72   &$-$0.38   &$-$0.47 \\
 000221.11+002149.3   & 00 02 21.1 &  00 21 49.6 & 3.07 &$-$28.0 & 12.9 & 16.8                  & 16.5                  &$-$0.03   &  0.21   &  0.15 \\
 001507.00$-$000800.9 & 00 15 07.0 &$-$00 08 01.3 & 1.70 &$-$27.1 & 12.5 & 12.6                  &  8.5                  &$-$0.73   &  0.00   &$-$0.20 \\
 073659.31+293938.4   & 07 36 59.3 &  29 39 38.4 & 2.46 &$-$26.6 &  9.9 &  1.0                  &  0.5                  &$-$1.29   &$-$1.83   &$-$1.73 \\
 074917.62+351633.8   & 07 49 17.6 &  35 16 33.9 & 2.27 &$-$26.5 & 40.8 & 36.9                  & 34.9                  &$-$0.10   &$-$0.08   &$-$0.08 \\
 080220.51+303543.0   & 08 02 20.5 &  30 35 43.0 & 1.64 &$-$26.8 & 64.4 & 30.6                  & 27.3                  &$-$0.21   &$-$0.59   &$-$0.49 \\
 080248.43+291734.2   & 08 02 48.4 &  29 17 34.3 & 2.38 &$-$27.0 &195.2 & 65.2                  & 41.4                  &$-$0.84   &$-$0.87   &$-$0.81 \\
 080415.80+300430.8   & 08 04 15.8 &  30 04 30.6 & 2.11 &$-$25.6 & 29.9 & 10.9                  &  6.4                  &$-$0.99   &$-$0.81   &$-$0.81 \\
 081520.94+323512.9   & 08 15 20.9 &  32 35 12.9 & 1.58 &$-$27.2 &  8.5 &  5.9                  &  5.0                  &$-$0.31   &$-$0.29   &$-$0.30 \\
 083301.64+353133.8   & 08 33 01.6 &  35 31 33.6 & 2.72 &$-$26.8 & 83.6 & 33.1                  & 24.1                  &$-$0.59   &$-$0.74   &$-$0.78 \\
 083629.57+345544.8   & 08 36 29.5 &  34 55 45.0 & 1.76 &$-$27.3 & 10.2 &  7.0                  &  6.3                  &$-$0.19   &$-$0.30   &$-$0.31 \\
 083757.91+383727.1   & 08 37 57.9 &  38 37 27.1 & 2.05 &$-$26.6 &  8.8 &  8.3                  &  7.7                  &$-$0.14   &$-$0.05   &$-$0.06 \\
 084617.52+375718.7   & 08 46 17.5 &  37 57 18.7 & 2.58 &$-$26.6 & 18.6 & 12.4\tablenotemark{b} & 29.2\tablenotemark{b} &  1.59   &$-$0.32   &  0.31 \\
 090745.46+382739.0   & 09 07 45.5 &  38 27 39.0 & 1.74 &$-$27.5 &160.5 & 47.9\tablenotemark{b} & 26.6\tablenotemark{b} &$-$1.09   &$-$0.96   &$-$0.97 \\
 090956.78+332357.5   & 09 09 56.8 &  33 23 57.7 & 1.49 &$-$25.8 & 35.1 & 29.9                  & 30.8                  &  0.06   &$-$0.13   &$-$0.07 \\
 091054.17+375915.0   & 09 10 54.1 &  37 59 15.1 & 2.16 &$-$28.9 &265.9 &  137.0              &  105.0              &$-$0.49   &$-$0.53   &$-$0.52 \\
 091639.78+390833.8   & 09 16 39.7 &  39 08 33.5 & 1.88 &$-$26.5 & 17.2 & 11.8                  & 11.2                  &$-$0.10   &$-$0.30   &$-$0.23 \\
 092156.27+305157.1   & 09 21 56.2 &  30 51 57.3 & 3.06 &$-$28.2 & 15.6 &  8.9                  &  8.1                  &$-$0.17   &$-$0.45   &$-$0.38 \\
 092726.55+383629.0   & 09 27 26.5 &  38 36 29.0 & 2.62 &$-$27.5 & 49.9 & 41.9                  & 34.2                  &$-$0.38   &$-$0.14   &$-$0.21 \\
 093152.76+343920.7   & 09 31 52.7 &  34 39 21.1 & 2.31 &$-$27.0 & 18.2 & 17.5                  & 18.4                  &  0.09   &$-$0.03   &$-$0.03 \\
 095537.94+333503.9   & 09 55 37.9 &  33 35 03.9 & 2.49 &$-$29.3 & 36.6 & 79.2                  & 56.2                  &$-$0.64   &  0.62   &  0.18 \\
 095555.68+351652.7   & 09 55 55.6 &  35 16 52.8 & 1.62 &$-$27.4 &  9.7 & 15.2                  & 14.7                  &$-$0.06   &  0.36   &  0.21 \\
 095733.45+384753.3   & 09 57 33.4 &  38 47 53.4 & 2.46 &$-$26.5 & 35.3 & 13.2                  & 10.4                  &$-$0.44   &$-$0.78   &$-$0.69 \\
 100146.19+371711.5   & 10 01 46.2 &  37 17 11.9 & 1.73 &$-$27.2 & 11.5 & 11.4                  &  8.3                  &$-$0.59   &$-$0.01   &$-$0.19 \\
 100302.60+404813.1   & 10 03 02.5 &  40 48 13.0 & 2.37 &$-$27.4 & 24.9 & 13.2                  &  8.2                  &$-$0.88   &$-$0.51   &$-$0.58 \\
 100929.99+321848.8   & 10 09 29.9 &  32 18 48.8 & 2.10 &$-$26.3 & 10.6 &  7.0                  &  4.7                  &$-$0.74   &$-$0.33   &$-$0.43 \\
 103510.99+351019.6   & 10 35 10.9 &  35 10 19.5 & 1.96 &$-$28.0 & 22.8 & 41.4                  & 53.2                  &  0.46   &  0.47   &  0.47 \\
 103551.16+375641.7   & 10 35 51.1 &  37 56 41.5 & 1.51 &$-$28.1 & 54.2 & 53.9                  & 54.5                  &  0.02   &$-$0.01   &  0.00 \\
 103648.50+370307.5   & 10 36 48.4 &  37 03 07.7 & 1.60 &$-$26.9 & 26.7 & 42.1                  & 50.4                  &  0.33   &  0.36   &  0.35 \\
 104217.80+373931.7   & 10 42 17.8 &  37 39 31.5 & 1.67 &$-$26.1 &  7.4 &  6.3                  &  5.6                  &$-$0.22   &$-$0.14   &$-$0.15 \\
 104229.18+381111.2   & 10 42 29.1 &  38 11 11.3 & 2.63 &$-$28.5 & 11.8 & 55.0                  & 81.5                  &  0.73   &  1.23   &  0.82 \\
 104713.15+353115.6   & 10 47 13.1 &  35 31 15.4 & 2.44 &$-$26.9 & 28.6 & 29.1                  & 23.7                  &$-$0.38   &  0.01   &$-$0.08 \\
 105611.77+315616.5   & 10 56 11.7 &  31 56 16.3 & 1.51 &$-$26.4 & 11.1 & 14.7                  & 15.5                  &  0.10   &  0.22   &  0.15 \\
 112403.46+381523.8   & 11 24 03.4 &  38 15 23.8 & 2.36 &$-$26.4 & 22.8 & 13.9\tablenotemark{a} &  6.2\tablenotemark{a} &$-$1.50   &$-$0.40   &$-$0.51 \\
 113442.06+411329.8   & 11 34 42.0 &  41 13 30.1 & 1.69 &$-$27.7 & 75.0 & 33.2                  &   27.0                  &$-$0.38   &$-$0.65   &$-$0.58 \\
 113854.51+394553.6   & 11 38 54.5 &  39 45 53.6 & 2.16 &$-$27.5 & 21.2 & 28.4                  & 34.9                  &  0.38   &  0.23   &  0.29 \\
 114023.06+372815.1   & 11 40 23.0 &  37 28 15.2 & 2.58 &$-$28.3 & 16.1 &  9.4                  &  6.6                  &$-$0.66   &$-$0.43   &$-$0.48 \\
 115019.60+334144.8   & 11 50 19.5 &  33 41 43.8 & 1.49 &$-$26.7 &229.8 &  101.0              & 63.9                  &$-$0.85   &$-$0.66   &$-$0.76 \\
 120003.89+410842.3   & 12 00 03.8 &  41 08 42.6 & 1.86 &$-$27.1 &162.8 & 52.3                  & 33.8                  &$-$0.81   &$-$0.91   &$-$0.85 \\
 120943.34+393642.6   & 12 09 43.3 &  39 36 42.7 & 2.33 &$-$27.2 & 23.8 & \nodata               & 22.4                  & \nodata & \nodata &$-$0.04 \\
 121303.80+324736.8   & 12 13 03.8 &  32 47 36.9 & 2.51 &$-$27.7 &133.6 & 65.5                  &  102.0                  &  0.83   &$-$0.57   &$-$0.60 \\
 124206.57+370138.9   & 12 42 06.6 &  37 01 38.4 & 2.71 &$-$27.3 & 26.0 &  8.6                  &  5.8                  &$-$0.73   &$-$0.88   &$-$0.86 \\
 124402.01+401642.3   & 12 44 01.9 &  40 16 42.4 & 2.29 &$-$27.4 & 70.4 & 26.7                  & 16.8                  &$-$0.86   &$-$0.77   &$-$0.79 \\
 130248.13+393002.3   & 13 02 48.1 &  39 30 02.0 & 2.44 &$-$27.8 &100.1 & 40.2                  & 26.2                  &$-$0.79   &$-$0.73   &$-$0.76 \\
 130941.51+404757.2   & 13 09 41.4 &  40 47 57.3 & 2.91 &$-$27.9 & 38.6 &  123.0                &  100.0               &$-$0.38   &  0.93   &  0.51 \\
 131016.75+334419.7   & 13 10 16.7 &  33 44 19.6 & 1.79 &$-$28.2 & 17.7 & 30.4                  & 28.7                  &$-$0.11   &  0.43   &  0.27 \\
 133139.90+312757.7   & 13 31 39.9 &  31 27 57.4 & 3.02 &$-$28.4 & 34.6 & 13.6                  &  9.7                  &$-$0.63   &$-$0.74   &$-$0.72 \\
 133403.83+370104.0   & 13 34 03.8 &  37 01 04.0 & 1.78 &$-$28.0 & 13.3 & 11.5                  &  7.8                  &$-$0.72   &$-$0.12   &$-$0.16 \\
 133410.53+384643.2   & 13 34 10.5 &  38 46 43.1 & 1.72 &$-$26.4 & 73.6 &  114.0                  & 91.5                  &$-$0.42   &  0.35   &  0.19 \\
 134103.71+321429.8   & 13 41 03.7 &  32 14 30.0 & 2.15 &$-$27.0 & 42.1 & 28.8                  & 20.3                  &$-$0.65   &$-$0.30   &$-$0.39 \\
 134253.64+390223.6   & 13 42 53.6 &  39 02 23.6 & 1.72 &$-$26.5 & 13.4 & 12.1                  & \nodata               & \nodata &$-$0.08   &$-$0.08 \\
 134804.34+284025.3   & 13 48 04.3 &  28 40 25.4 & 2.45 &$-$28.5 & 78.0 & 76.2                  & 74.4                  &$-$0.04   &$-$0.02   &$-$0.03 \\
 135819.85+321926.2   & 13 58 19.8 &  32 19 26.5 & 2.83 &$-$28.0 & 20.4 & 29.2                  & 28.1                  &$-$0.07   &  0.28   &  0.09 \\
 140911.40+354735.3   & 14 09 11.3 &  35 47 35.4 & 1.74 &$-$26.5 &  8.2 & 10.6                  &  9.1                  &$-$0.28   &  0.20   &  0.11 \\
 141536.91+340441.8   & 14 15 36.9 &  34 04 41.9 & 2.48 &$-$26.5 & 16.1 & 8.5                  & 7.7                  &$-$0.17   &$-$0.51   &$-$0.48 \\
 142009.33+392738.5   & 14 20 09.3 &  39 27 38.7 & 2.29 &$-$27.5 & 38.3 & 20.8                  & 16.2                  &$-$0.46   &$-$0.49   &$-$0.48 \\
 142522.79+394337.9   & 14 25 22.8 &  39 43 38.2 & 3.36 &$-$27.5 & 26.9 & 11.0                  &  9.0                  &$-$0.37   &$-$0.71   &$-$0.61 \\
 145924.24+340113.1   & 14 59 24.2 &  34 01 13.1 & 2.79 &$-$28.9 & 21.7 & 67.8                  & 63.7                  &$-$0.12   &  0.91   &  0.27 \\
 150149.45+340831.6   & 15 01 49.4 &  34 08 31.7 & 3.36 &$-$27.6 & 18.9 & 14.0                  & 12.0                  &$-$0.29   &$-$0.24   &$-$0.25 \\
 150545.05+411726.8   & 15 05 45.0 &  41 17 26.8 & 1.52 &$-$27.2 & 34.2 & 18.4                  & 13.1                  &$-$0.63   &$-$0.49   &$-$0.53 \\
 150937.01+401713.8   & 15 09 37.0 &  40 17 13.6 & 2.32 &$-$27.4 & 15.4 & 34.2                  & 29.9                  &$-$0.25   &  0.64   &  0.41 \\
 151640.38+343833.2   & 15 16 40.3 &  34 38 33.2 & 1.47 &$-$25.0 & 39.0 & 22.5                  & 15.2                  &$-$0.73   &$-$0.44   &$-$0.49 \\
 152314.88+381402.1   & 15 23 14.8 &  38 14 02.1 & 3.16 &$-$28.0 & 47.7 & 40.2                  & 28.2                  &$-$0.66   &$-$0.14   &$-$0.27 \\
 154042.97+413816.3   & 15 40 42.9 &  41 38 16.6 & 2.52 &$-$29.2 & 18.2 & 75.1                  & 88.3                  &  0.30   &  1.13   &  0.85 \\
 154135.85+374051.1   & 15 41 35.8 &  37 40 50.9 & 2.25 &$-$26.3 & 98.6 &  136.0              &  108.0                 &$-$0.43   &  0.26   &  0.18 \\
 154551.06+393643.7   & 15 45 51.0 &  39 36 43.5 & 1.75 &$-$25.8 &  9.8 & 19.6                  & 14.0                  &$-$0.62   &  0.55   &  0.29 \\
 154644.24+311711.3   & 15 46 44.2 &  31 17 11.4 & 2.12 &$-$26.6 & 12.6 & 12.5                  &  9.4                  &$-$0.52   &$-$0.01   &$-$0.12 \\
 155315.74+403926.8   & 15 53 15.7 &  40 39 26.7 & 2.75 &$-$28.2 & 50.4 & 48.9                  & 26.0                  &$-$1.17   &$-$0.02   &$-$0.15 \\
 161754.55+363917.4   & 16 17 54.5 &  36 39 17.7 & 2.29 &$-$26.2 & 10.7 &  9.5                  &  7.6                  &$-$0.41   &$-$0.10   &$-$0.16 \\
 161948.58+382729.9   & 16 19 48.5 &  38 27 30.0 & 2.07 &$-$27.2 & 17.0 &  5.1                  &  4.2                  &$-$0.36   &$-$0.96   &$-$0.80 \\
 162708.50+301530.2   & 16 27 08.5 &  30 15 30.2 & 2.14 &$-$26.4 & 11.6 &  6.0                  &  5.3                  &$-$0.23   &$-$0.53   &$-$0.46 \\
 165137.52+400219.0   & 16 51 37.5 &  40 02 18.7 & 2.34 &$-$29.1 & 43.9 & 53.5                  & 39.2                  &$-$0.58   &  0.16   &  0.13 \\
 165559.40+404811.0   & 16 55 59.4 &  40 48 11.1 & 1.83 &$-$26.5 & 25.8 & 20.2                  & 17.2                  &$-$0.30   &$-$0.19   &$-$0.23 \\
 165802.61+360504.1   & 16 58 02.6 &  36 05 04.3 & 2.12 &$-$28.1 & 64.8 & 54.9                  & 51.9                  &$-$0.10   &$-$0.13   &$-$0.13 \\    
 \enddata
 \tablecomments{Properties of the non-BAL quasar sample.  Columns are the same as Table~\ref{balpropstbl}, with the BAL type column omitted.}
 \tablenotetext{a}{Observations not simultaneous- 8.4 GHz observed with VLA in 2010, 4.9 GHz observed with EVLA in 2011}
 \tablenotetext{b}{Observations not simultaneous- 4.9 GHz observed with VLA in 2010, 8.4 GHz observed with EVLA in 2011}

\end{deluxetable}

\clearpage

\begin{deluxetable}{rcccc}
 \tabletypesize{\scriptsize}
 \tablewidth{0pt}
 \tablecaption{Radio spectral index statistics\label{alphastatstbl}}
 \tablehead{
   \colhead{Measurement} & \colhead{$n$} & \colhead{Mean} & \colhead{Median} & \colhead{$\sigma$}
  }
   \startdata
BAL $\alpha_{8.4}^{4.9}$         & 72 &$-$0.73 &$-$0.69 & 0.55 \\[3pt]
BAL $\alpha_{4.9}^{1.4}$         & 73 &$-$0.44 &$-$0.50 & 0.44 \\[3pt]
BAL $\alpha_{fit}$                      & 73 &$-$0.50 &$-$0.49 & 0.38 \\[3pt]
non-BAL $\alpha_{8.4}^{4.9}$ & 72 &$-$0.36 &$-$0.38 & 0.49 \\[3pt]
non-BAL $\alpha_{4.9}^{1.4}$ & 73 &$-$0.18 &$-$0.14 & 0.54 \\[3pt]
non-BAL $\alpha_{fit}$              & 74 &$-$0.23 &$-$0.20 & 0.45 \\[3pt]
\hline\\
\hline\\
cBAL $\alpha_{8.4}^{4.9}$         & 63 &$-$0.66 &$-$0.58 & 0.55 \\[3pt]
cBAL $\alpha_{4.9}^{1.4}$         & 63 &$-$0.38 &$-$0.45 & 0.44 \\[3pt]
cBAL $\alpha_{fit}$			 & 63 &$-$0.44 &$-$0.45 & 0.37 \\[3pt]
cnon-BAL $\alpha_{8.4}^{4.9}$ & 56 &$-$0.30 &$-$0.31 & 0.52 \\[3pt]
cnon-BAL $\alpha_{4.9}^{1.4}$ & 57 &$-$0.07 &$-$0.10 & 0.53 \\[3pt]
cnon-BAL $\alpha_{fit}$		 & 58 &$-$0.13 &$-$0.12 & 0.44 \\[3pt]
   \enddata
 \tablecomments{The top half of the table gives the mean, median, and standard deviation ($\sigma$) of the various spectral index measurements for the full BAL and normal quasar samples.  The number of objects with each measurement is given in the column labeled $n$.  The second half shows the statistics restricting the samples to only compact sources, defined as having $\Theta < 0.1$ (see text).}
\end{deluxetable}

\begin{deluxetable}{ccccccc}
 \tabletypesize{\scriptsize}
 \tablewidth{0pt}
 \tablecaption{Statistical tests on $\alpha$ distributions\label{stattesttbl}}
 \tablehead{
   \colhead{Measurement} & \colhead{$n$ BAL} & \colhead{$n$ non-BAL} & \colhead{$D_{ks}$} &  \colhead{$P_{ks}$} & \colhead{$Z_{rs}$}  &  \colhead{$P_{rs}$}
  }
   \startdata
$\alpha_{8.4}^{4.9}$         & 72 & 72 & 0.347 & 0.0002 & 4.00 & $3.1\times 10^{-5}$ \\[3pt]
$\alpha_{4.9}^{1.4}$         & 73 & 73 & 0.287 & 0.0036 & 3.18 & 0.0007                        \\[3pt]
$\alpha_{fit}$                      & 73 & 74 & 0.322 & 0.0007 & 3.76 & $8.4 \times 10^{-5}$ \\[3pt]
\hline \\
\hline \\
c $\alpha_{8.4}^{4.9}$      & 63 & 56 & 0.337 & 0.0016 & 3.63 & 0.0001                        \\[3pt]
c $\alpha_{4.9}^{1.4}$      & 63 & 57 & 0.342 & 0.0012 & 3.70 & 0.00011                      \\[3pt]
c $\alpha_{fit}$		         & 63 & 58 & 0.394 & 0.0001 & 4.19 & $1.4 \times 10^{-5}$ \\[3pt]
   \enddata
 \tablecomments{The top half of the table shows the K-S test statistic ($D_{ks}$) followed by the corresponding p-value ($P_{ks}$), and the mean rank-sum test statistic ($Z_{rs}$) with its corresponding p-value ($P_{rs}$) on the BAL and normal quasar distributions for three measurements of the spectral index.  The bottom half shows the results for these indices restricted to compact sources.}
\end{deluxetable}

\clearpage

\begin{deluxetable}{ccccc}
 \tabletypesize{\scriptsize}
 \tablewidth{0pt}
 \tablecaption{Variable sources at 1.4 GHz\label{vartbl}}
 \tablehead{
   \colhead{Source (SDSS J)} & \colhead{$S_1$ (FIRST)} & \colhead{$S_2$ (NVSS)} & \colhead{$Var_{\Delta S}$} & \colhead{$\sigma_{var}$}
  }
   \startdata
125243.85+005320.1   &  10.3   &   15.4   &  0.50   &   4.56   \\
135550.30+361627.6   &  10.7   &   15.3   &  0.43   &   5.74   \\
\hline
093152.76+343920.7   &  18.3   &   25.7   &  0.41   &   4.94   \\       
104229.18+381111.2   &  11.8   &   6.7     &  0.76   &   7.15   \\
140911.40+354735.3   &  8.2     &   12.2   &  0.49   &   4.03   \\
161754.55+363917.4   &  10.7   &   14.6   &  0.36   &   4.81   \\
   \enddata
 \tablecomments{These six sources may have significant variability, based on $\sigma_{var}$.  The two above the horizontal line are from the BAL sample, and the four below it are from the unabsorbed sample.}
\end{deluxetable}

\clearpage

\begin{figure}
 \centering
  \figurenum{2}
   \subfloat[][Fig. 2$a$]{\includegraphics[width=5.2in]{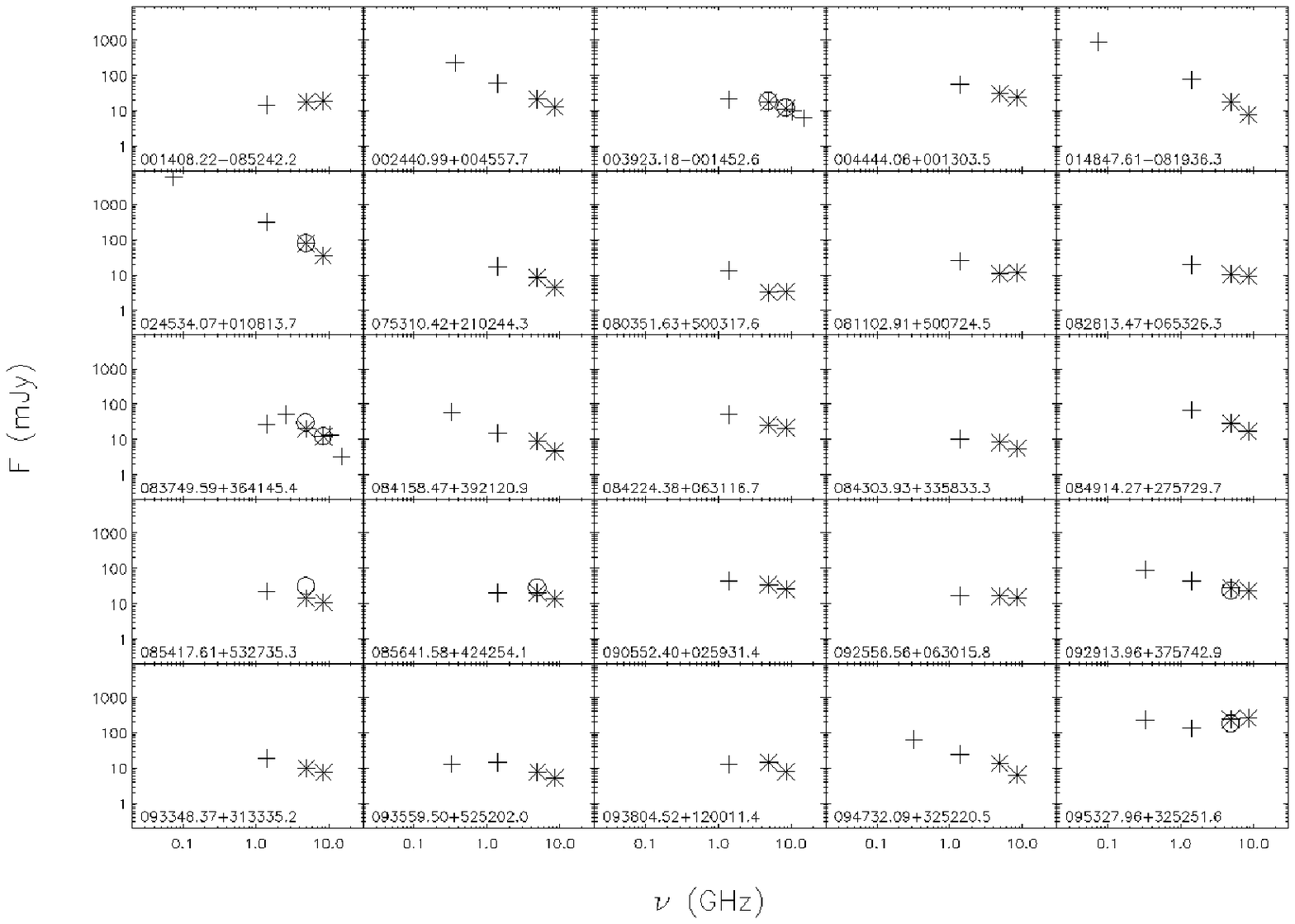}}
    \caption{Radio spectra of the BAL sample, including literature data (see text for details).\label{balspectra}}
\end{figure}

\begin{figure}
 \ContinuedFloat
 \centering
  \subfloat[][Fig. 2$b$]{\includegraphics[width=5.2in]{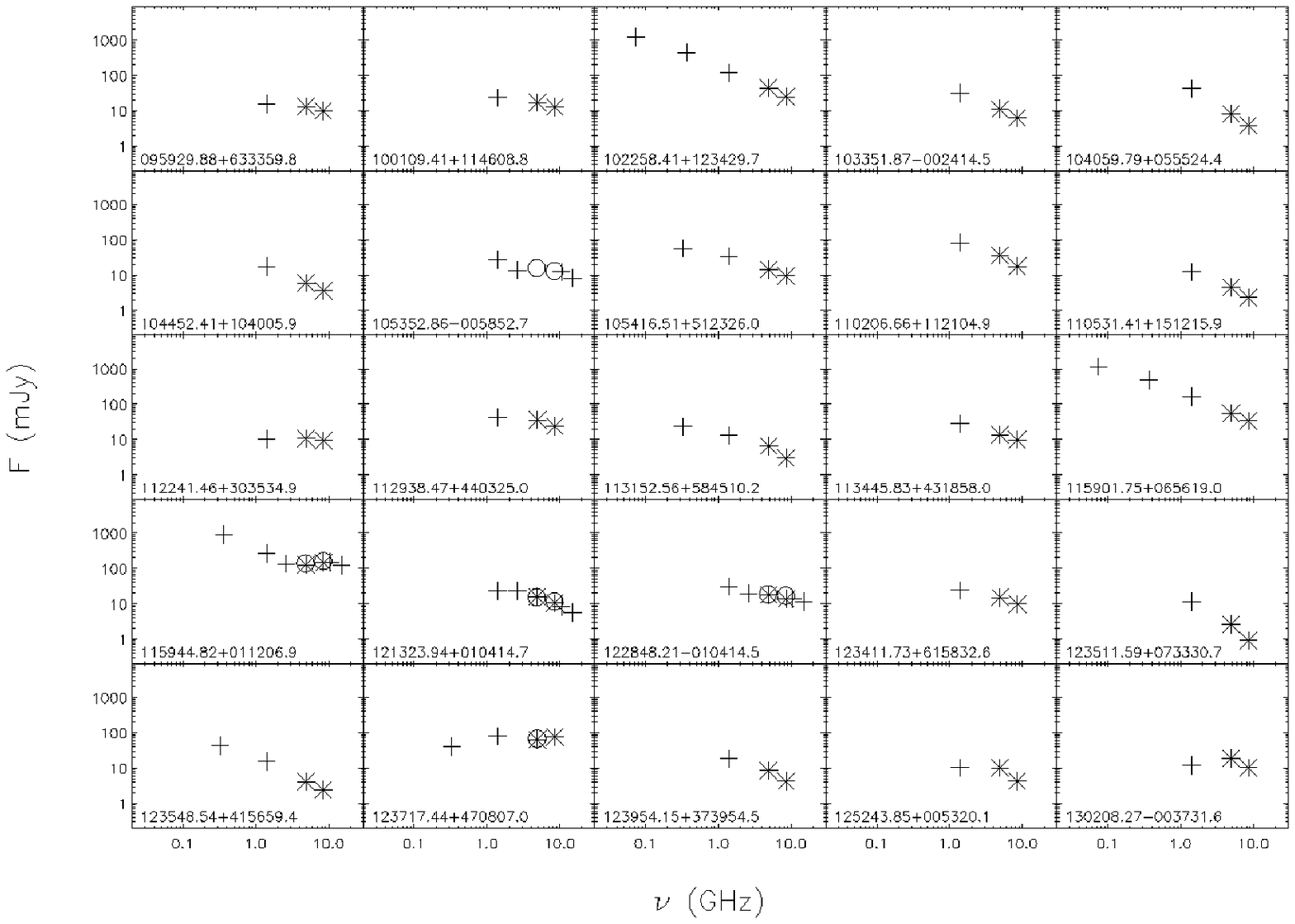}}
\end{figure}

\begin{figure}
 \ContinuedFloat
 \centering
  \subfloat[][Fig. 2$c$]{\includegraphics[width=5.2in]{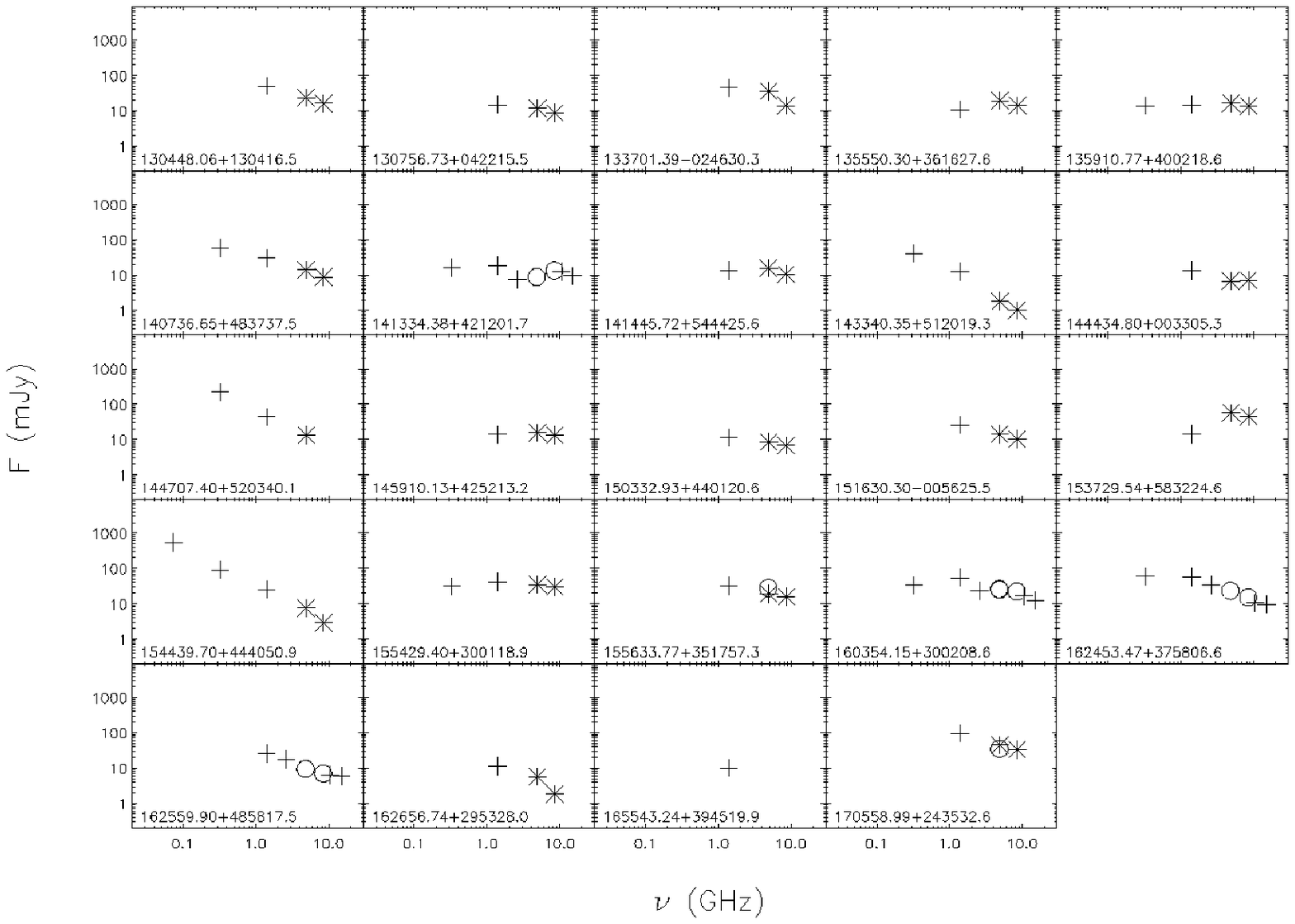}}
\end{figure}

\begin{figure}
 \centering
  \figurenum{3}
   \subfloat[][Fig. 3$a$]{\includegraphics[width=5in]{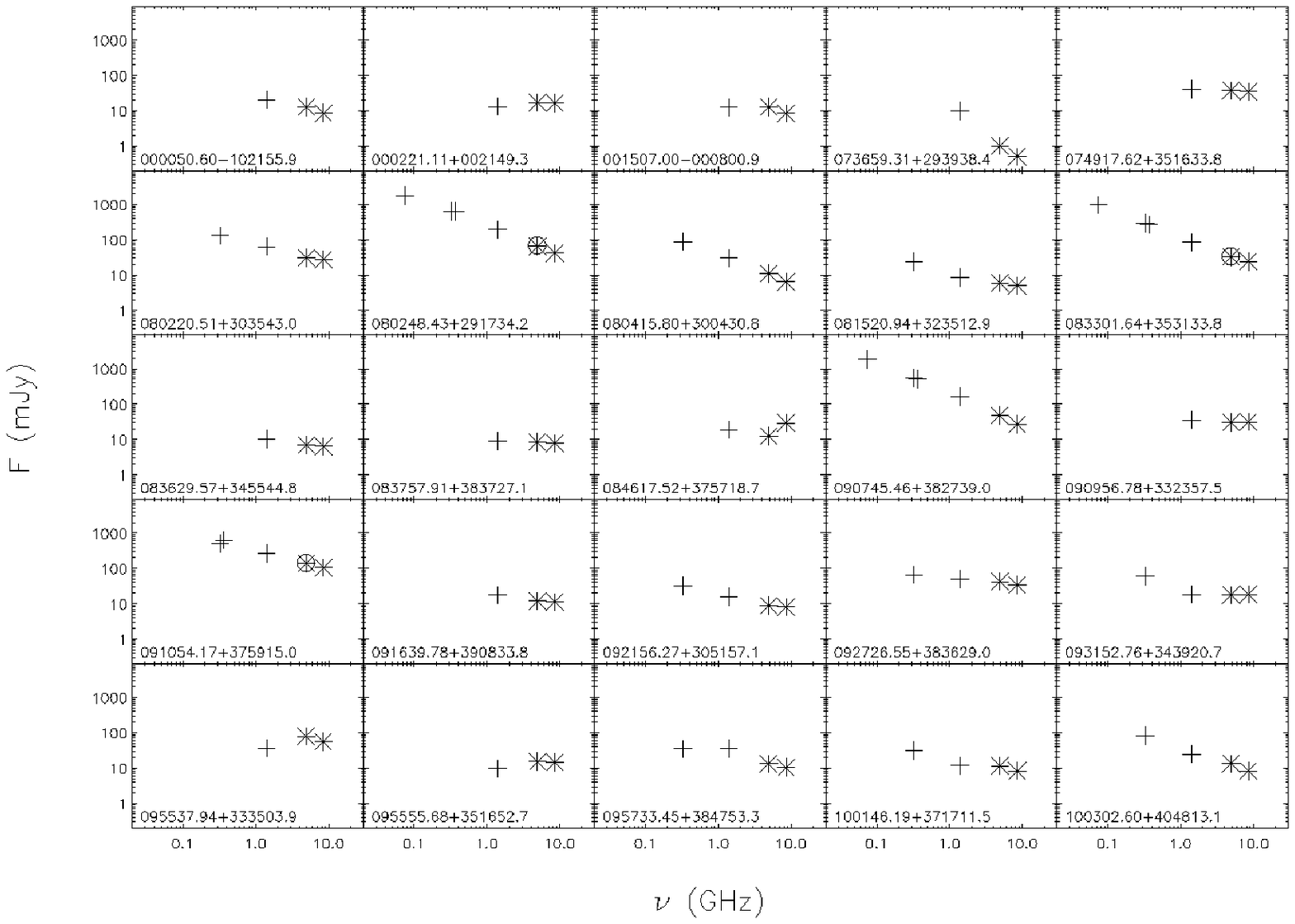}}
    \caption{Radio spectra of the normal quasar sample, including literature values (see text for details).\label{nbalspectra}}
\end{figure}

\begin{figure}
 \ContinuedFloat
 \centering
  \subfloat[][Fig. 3$b$]{\includegraphics[width=5in]{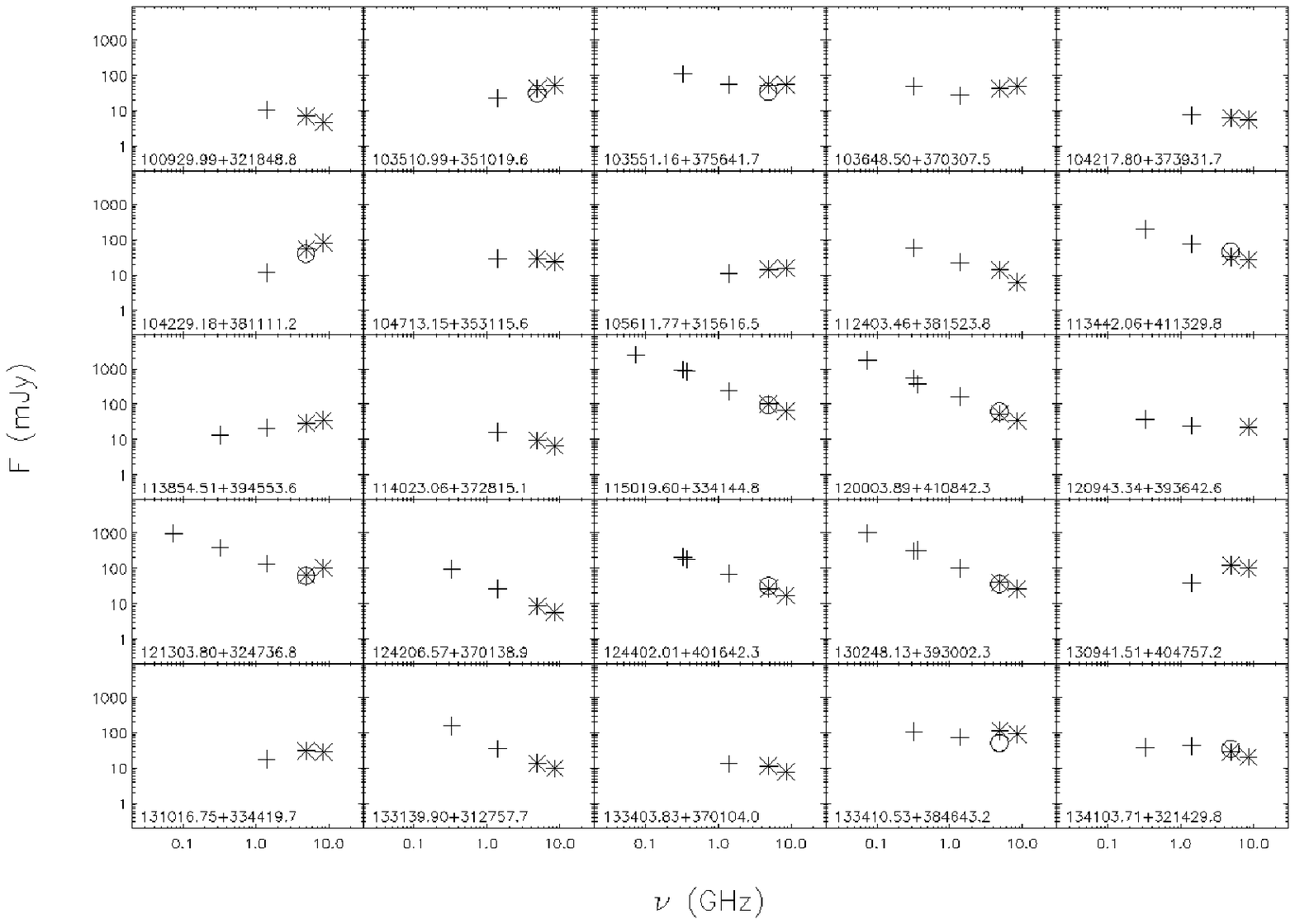}}
\end{figure}

\begin{figure}
 \ContinuedFloat
 \centering
  \subfloat[][Fig. 3$c$]{\includegraphics[width=5in]{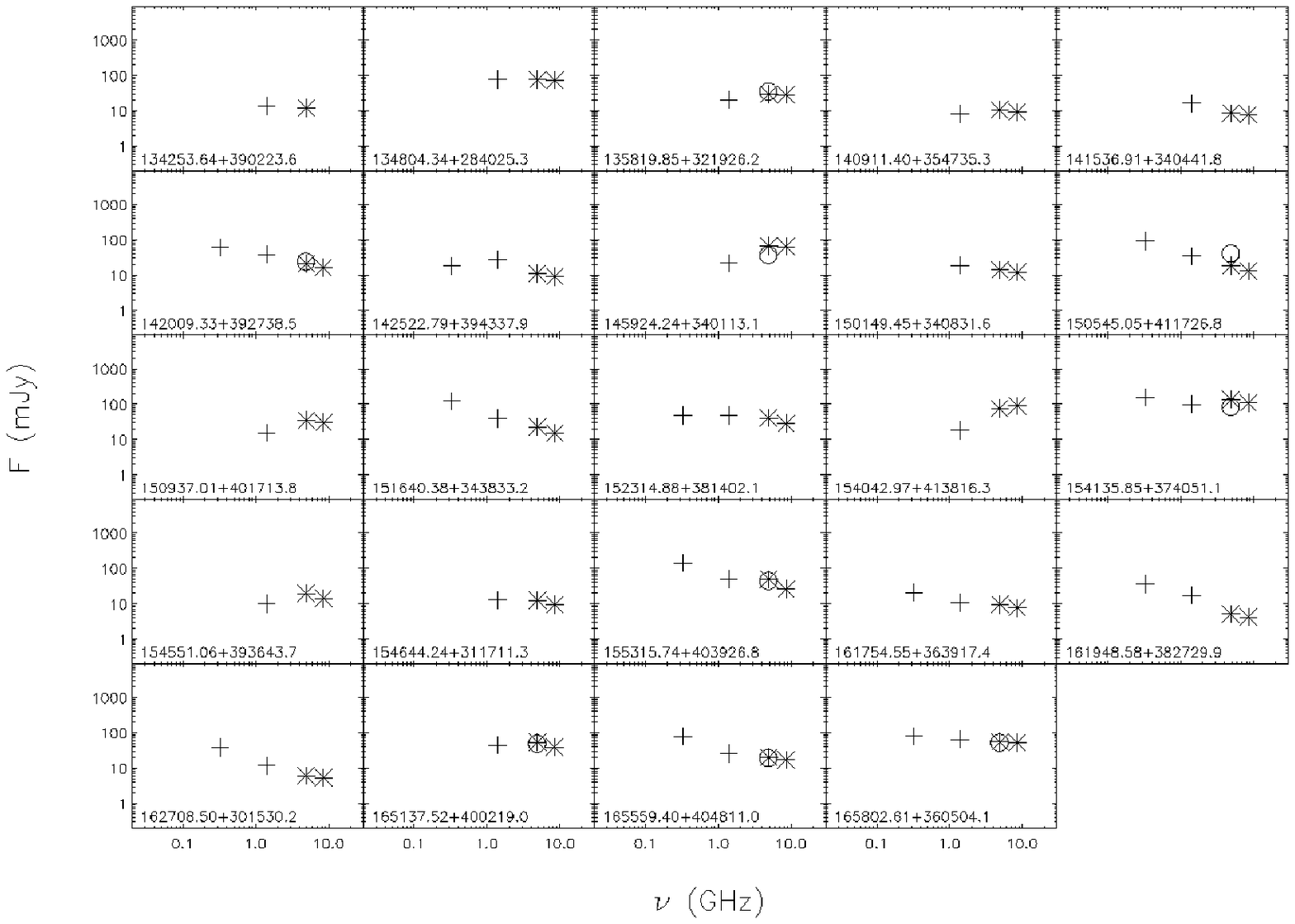}}
\end{figure}

\clearpage

\end{document}